%% file: SUS-17-010_temp.tex
\pdfoutput=1

\documentclass[11pt,twoside,a4paper,cmspaper,final,collab]{cms-tdr}
\def\svnVersion{482223P}\def\cmsCernNoTag{CERN-EP-2018-186}\def\cmsCernDate{\today}\def\cmsMessage{Published in the Journal of High Energy Physics as \href{http://dx.doi.org/10.1007/JHEP11(2018)079}{\doi{10.1007/JHEP11(2018)079.}}}

\begin{document}\cmsNoteHeader{SUS-17-010}

\hyphenation{had-ron-i-za-tion}
\hyphenation{cal-or-i-me-ter}
\hyphenation{de-vices}
\RCS$HeadURL: svn+ssh://svn.cern.ch/reps/tdr2/papers/SUS-17-010/trunk/SUS-17-010.tex $
\RCS$Id: SUS-17-010.tex 481952 2018-11-21 12:00:49Z bchazinq $

\newlength\cmsFigWidth
\providecommand{\cmsTable}[1]{\resizebox{\textwidth}{!}{#1}}
\newlength\cmsTabSkip\setlength{\cmsTabSkip}{1ex}

\newcommand{\FastSim} {{\textsc{FastSim}}\xspace}
\newcommand{\lhcE}[1]{\ensuremath{\sqrt{s}={#1}\TeV}}
\newcommand{\Lum}{35.9\fbinv}
\newcommand{\invM}[1]{\ensuremath{m_{#1}}\xspace}
\newcommand{\mll}{\invM{\ell\ell}}
\newcommand{\MT}[1]{\ensuremath{m_\mathrm{#1}}\xspace}
\newcommand{\mtll}{\ensuremath{\MT{T2}(\ell\ell)}\xspace}
\newcommand{\tvec}[1]{\ensuremath{{\vec p}^{\,\text{#1}}_{\mathrm{T}}}}
\newcommand{\sell}{\ensuremath{\widetilde{\ell}}}
\newcommand{\mumu}{\ensuremath{\Pgm\Pgm}\xspace}
\newcommand{\ee}{\ensuremath{\Pe\Pe}\xspace}
\newcommand{\emu}{\ensuremath{\Pe\Pgm}\xspace}
\newcommand{\pp}{\ensuremath{\Pp\Pp}\xspace}
\newcommand{\tW}{\ensuremath{\PQt\PW}\xspace}
\newcommand{\WW}{\ensuremath{\PW\PW}\xspace}
\newcommand{\WZ}{\ensuremath{\PW\PZ}\xspace}
\newcommand{\ZZ}{\ensuremath{\PZ\PZ}\xspace}
\newcommand{\ttW}{\ensuremath{\ttbar\PW}\xspace}
\newcommand{\ttZ}{\ensuremath{\ttbar\PZ}\xspace}
\newcommand{\DY}{\text{Drell--Yan}\xspace}
\newcommand{\Dm}{\ensuremath{\Delta m}\xspace}
\newcommand{\Dphi}{\ensuremath{\Delta\phi}\xspace}
\newcommand{\DZmass}{\ensuremath{\abs{\mll - \invM{\PZ}}}}
\newcommand{\ptisr}{\ensuremath{\pt^\mathrm{ISR}}\xspace}
\newcommand{\N}[1]{\ensuremath{N_\text{#1}}\xspace}
\newcommand{\Njetisr}{\ensuremath{\N{jet}^\mathrm{ISR}}\xspace}
\newcommand{\Irel}{\ensuremath{I_\text{rel}}\xspace}
\newcommand{\dxy}{\ensuremath{d_\mathrm{0}}\xspace}
\newcommand{\dz}{\ensuremath{d_{z}}\xspace}
\newcommand{\Sd}{\ensuremath{S_\mathrm{3D}^\mathrm{d}}\xspace}
\newcommand{\Reg}[3]{\ensuremath{\text{#1}_\text{#2}^\text{#3}}\xspace}
\newcommand{\TChipmSlep}{\ensuremath{\PSGcpmDo\to\Pgn\sell\to\Pgn\ell\PSGczDo}\xspace}
\newcommand{\TChipmSneu}{\ensuremath{\PSGcpmDo\to\ell\sNu\to\ell\Pgn\PSGczDo}\xspace}
\newcommand{\TChipmDecay}{\ensuremath{\PSGcpmDo\to\sell\nu\text{(}\ell\sNu\text{)}\to\ell\nu\PSGczDo}}
\newcommand{\BF}{\ensuremath{\mathcal{B}}\xspace}

\cmsNoteHeader{SUS-17-010}
\title{Searches for pair production of charginos and top squarks in final states with two oppositely charged leptons in proton-proton collisions at \texorpdfstring{$\sqrt{s}= 13\TeV$}{sqrt(s)=13 TeV}}

\date{\today}

\abstract{A search for pair production of supersymmetric particles in events with two oppositely charged leptons (electrons or muons) and missing transverse momentum is reported. The data sample corresponds to an integrated luminosity of \Lum of proton-proton collisions at \lhcE{13} collected with the CMS detector during the 2016 data taking period at the LHC.
No significant deviation is observed from the predicted standard model background. The results are interpreted in terms of several simplified models for chargino and top squark pair production, assuming $R$-parity conservation and with the neutralino as the lightest supersymmetric particle. When the chargino is assumed to undergo a cascade decay through sleptons, with a slepton mass equal to the average  of the chargino and neutralino masses, exclusion limits at 95\% confidence level are set on the masses of the chargino and neutralino up to 800 and 320\GeV, respectively. For top squark pair production, the search focuses on models with a small mass difference between the top squark and the lightest neutralino. When the top squark decays into an off-shell top quark and a neutralino, the limits extend up to 420 and 360\GeV for the top squark and neutralino masses, respectively.}

\hypersetup{
pdfauthor={CMS Collaboration},
pdftitle={Searches for pair production of charginos and top squarks in final states with two oppositely charged leptons in proton-proton collisions at sqrt(s)=13 TeV},
pdfsubject={CMS},
pdfkeywords={CMS, physics, supersymmetry}}

\maketitle

\section{Introduction}\label{sec:introduction}

The standard model (SM) of particle physics has so far been able to describe a wide variety of phenomena with outstanding precision.
However, the SM does not address the hierarchy problem between the Higgs boson mass and the Planck scale~\cite{PhysRevD.13.3333,Hierarchy1}, and does not contain a dark matter candidate to explain cosmological observations~\cite{DarkMatter1,DarkMatter2,DarkMatter3}.
Supersymmetry~\cite{Ramond:1971gb,Golfand:1971iw,Neveu:1971rx,Volkov:1972jx,Wess:1973kz,Wess:1974tw,Fayet:1974pd,Nilles:1983ge,SUSYPrimer} is an extension of the SM that assigns a fermion (boson) superpartner to every SM boson (fermion).
This theory can solve the hierarchy problem since the large quantum loop corrections to the Higgs boson mass, due mainly to the top quark, can be largely canceled by the analogous corrections from the top quark superpartner~\cite{Hierarchy2,Hierarchy3,Hierarchy4}.
Moreover, if \emph{R}-parity~\cite{Farrar:1978xj} is conserved, the lightest supersymmetric particle (LSP) is stable and, if massive, provides a good candidate for dark matter.

This paper presents a search for supersymmetric particle production in final states with two oppositely charged (OC) leptons ($\ell$) and missing transverse momentum stemming from the two LSPs. Only electrons (\Pe) and muons (\Pgm) are considered.
The search targets two specific signal scenarios with chargino (\PSGcpmDo) and top squark (\stone) pair production, using data from proton-proton (\pp) collisions at \lhcE{13} collected by the CMS experiment~\cite{Chatrchyan:2008zzk} at the CERN LHC in 2016, and corresponding to an integrated luminosity of \Lum.

\begin{figure}[hbtp]
  \centering
    \includegraphics[width=0.4\textwidth]{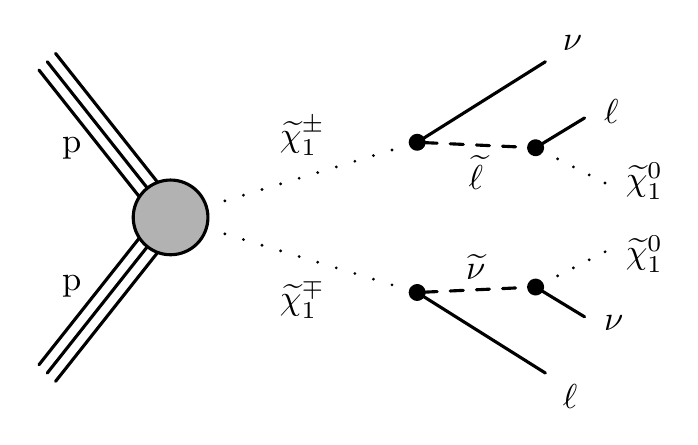}
    \includegraphics[width=0.4\textwidth]{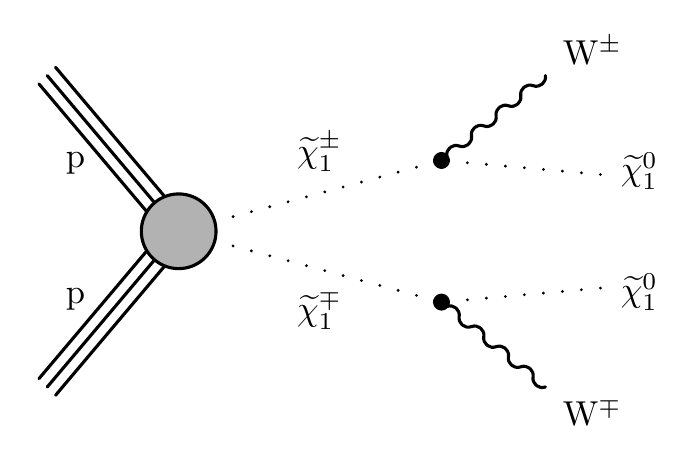}
    \caption{Simplified-model diagrams of chargino pair production with two benchmark decay modes: the left plot shows decays through intermediate sleptons or sneutrinos, while the right one displays prompt decays into a {\PW} boson and the lightest neutralino.}
    \label{Fig:TChipm}
\end{figure}

The results are interpreted in terms of simplified supersymmetric model spectra (SMS)~\cite{SMS1,SMS2,SMS3} scenarios.
The search for chargino pair production considers, as a reference, a model (Fig.~\ref{Fig:TChipm}, left)
where the charginos decay into a lepton, a neutrino (\Pgn), and the lightest neutralino (\PSGczDo) via an intermediate charged slepton (\TChipmSlep) or sneutrino (\TChipmSneu).
The three generations of sleptons are assumed to be degenerate, with a mass equal to the average of the
chargino and neutralino masses. The branching fractions ({\BF}'s) of the chargino decays into charged
sleptons or sneutrinos are assumed to be equal.
Results are also interpreted in terms of a second model (Fig.~\ref{Fig:TChipm}, right), where each chargino decays into the lightest neutralino and a {\PW} boson. Searches for chargino pair production have been previously published by the CMS Collaboration in the context of the former scenario using 8\TeV collision data~\cite{CMSTChipmSlepSnu} and by the ATLAS Collaboration in the context of both scenarios using 8\TeV~\cite{ATLASTChipmWW8TeV,ATLASTChipmStauSnu8TeV,ATLASTChipmSlepSnu8TeV} and 13\TeV~\cite{ATLASTChipmStauSnu13TeV,ATLASTChipmWW13TeV,ATLASTChipmSlepSnu13TeV} collision data.

\begin{figure}[hbtp]
  \centering
    \includegraphics[width=0.4\textwidth]{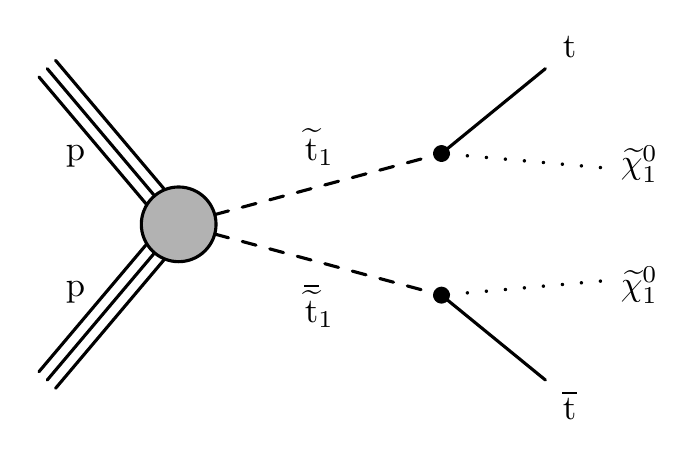}
    \includegraphics[width=0.4\textwidth]{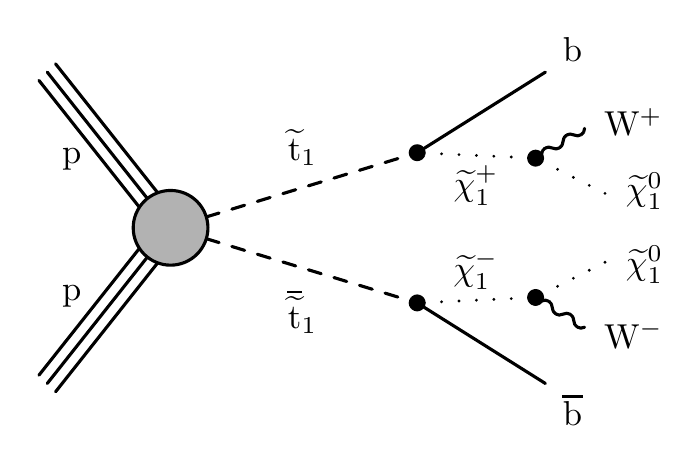}
    \caption{Simplified-model diagrams of top squark pair production with two benchmark decay modes of the top squark: the left plot shows decays into a top quark and the lightest neutralino, while the right one displays prompt decays into a bottom quark and a chargino, further decaying into a neutralino and a {\PW} boson.}
    \label{Fig:T2}
\end{figure}

The search for top squark pair production focuses on an SMS in which the top squark decays into a top quark and the lightest neutralino as shown in Fig.~\ref{Fig:T2} (left).
The analysis strategy is optimized for a compressed spectrum scenario where the mass difference (\Dm) between the top squark and the lightest neutralino lies between the top quark and {\PW} boson masses $\invM{\PW}<\Dm\lesssim\invM{\PQt}$. In this regime, the top quarks are produced off-shell, giving rise to final states with low-momentum bottom quarks which often fail to be identified.
Further interpretations of the results are given in terms of an additional model, where each of the pair-produced top squarks decays into a bottom quark and a chargino, which in turn decays into a {\PW} boson and the lightest neutralino, as shown in Fig.~\ref{Fig:T2} (right). In this model, the mass of the chargino is assumed to be equal to the average of the top squark and neutralino masses.
This work is complementary to another OC dilepton search published by the CMS Collaboration~\cite{ref:t2tt2lep}, aimed at testing models where $\Dm>\invM{\PQt}$, which result in signatures with on-shell top quarks and higher momentum particles.
With respect to that analysis, this search gains sensitivity in the compressed mass region by loosening the requirements on the jets from bottom quark hadronization and optimizing the signal event selection for the lower momentum carried by the neutralino LSPs.
The CMS Collaboration has also published other searches targeting the same signal models in the final states with exactly one lepton~\cite{ref:t2tt1lep} and with no leptons~\cite{ref:t2tt0lep}, with the latter also covering the four-body-decay of the top squark in the region $\Dm<80\GeV$. The ATLAS Collaboration published several searches addressing these signal models using all three final states~\cite{ref:ATLAS0lep,ref:ATLAS1lep,ATLASStop}.

The paper is organized as follows: Section~\ref{sec:detector} introduces the experimental apparatus;
Sections~\ref{sec:datasets} and~\ref{sec:reconstruction} describe the data and simulated event samples used in this search and the details on the reconstruction of the physics objects, respectively;
Section~\ref{sec:searchstrategy} presents the general strategy of the analysis;
Section~\ref{sec:bkgestimation} discusses the estimates of the contributions from SM processes to the selected events; Section~\ref{sec:systematics} details the sources of systematic uncertainties for signal and background processes;
Section~\ref{sec:results} reports the results and their interpretation in terms of the considered SMS;
and finally Section~\ref{sec:summary} summarizes the results of the search.

\section{The CMS detector}\label{sec:detector}

The central feature of the CMS apparatus is a superconducting solenoid of 6\unit{m} internal diameter, providing a magnetic field of 3.8\unit{T}.
In the inner part of the solenoid volume is a silicon pixel and strip tracker, which reconstructs the trajectories of the charged particles up to a pseudorapidity $\abs{\eta}<2.5$.
Outside the tracker, a lead tungstate crystal electromagnetic calorimeter (ECAL) and a brass and scintillator hadron calorimeter (HCAL), each composed of a barrel and two endcap sections, measure the energy of the particles in the region $\abs{\eta}<3$.
Forward calorimeters extend coverage provided by the barrel and endcap detectors up to $\abs{\eta}<5$.
The information from the tracker and calorimeter systems is merged to reconstruct electrons and hadronic jets.
Muons are detected in gas-ionization chambers embedded in the steel flux-return yoke outside the solenoid, covering the region $\abs{\eta}<2.4$.
The detector is nearly hermetic, allowing for momentum balance measurements in the plane transverse to the beam direction.
A more detailed description of the CMS detector, together with a definition of the coordinate system used and the relevant kinematic variables, can be found in Ref.~\cite{Chatrchyan:2008zzk}.

\section{Data and simulated samples}\label{sec:datasets}

Events of interest are selected using triggers~\cite{Khachatryan:2016bia} which require the presence of two leptons (\ee, \mumu, \emu).
The threshold on the transverse momentum (\pt) of the leading lepton is 23\GeV for the \ee and \emu triggers, and 17\GeV for the \mumu triggers. The threshold for the trailing lepton is 8 (12)\GeV for muons (electrons).
To increase the efficiency of the trigger selection, events are also accepted by triggers requiring at least one electron (muon) with $\pt>25$ (24)\GeV, passing tighter identification criteria than the ones applied in the double-lepton triggers.
The trigger performances are measured with leptons from $\PZ\to\ell^+\ell^-$ decays.
The combined efficiency of the dilepton and single-lepton triggers for signal events is found to range between 90 and 99\%, depending on the \pt and $\eta$ of the leptons.

Samples of Monte Carlo (MC) simulated events are used to study the contribution of SM processes to the selected data set and the expected acceptance for the different signal models. Events from top quark-antiquark pair (\ttbar) production are generated with \POWHEG v2~\cite{bib:powhegv2a,bib:powhegv2b,bib:powhegv2c} and normalized to the expected cross section calculated at next-to-next-to-leading order (NNLO) in perturbative quantum chromodynamics (QCD), including resummation of next-to-next-to-leading logarithmic (NNLL) soft gluon terms~\cite{Czakon:2011xx}.
Events with a single top quark produced in association with a {\PW} boson (\tW) are generated with \POWHEG v1~\cite{bib:powheg3} and normalized to an approximate NNLO cross section calculation~\cite{bib:tWxsec}.
Diboson production (\WW, \WZ, and \ZZ) via quark-antiquark annihilation is simulated at next-to-leading order (NLO) using \POWHEG v2~\cite{bib:powhegVVa,bib:powhegVVb}.
The yields of events from \WW production are scaled to the NNLO cross section~\cite{bib:WWxsec}.
Events from $\Pq\Paq\to\PZ\PZ$ production are reweighted via NNLO/NLO \emph{K} factors, as functions of the generated \ZZ system mass~\cite{HIG16041}.
Two additional sets of \emph{K} factors, as functions of the generated \ZZ system \pt and of the azimuthal separation (\Dphi) between the \PZ bosons, are used to evaluate the uncertainty in the kinematic properties of \ZZ production.
Diboson production via gluon fusion is simulated using \MCFM v7~\cite{bib:MCFM}, and LO cross sections obtained from the generator are corrected with the NNLO/LO \emph{K} factors~\cite{bib:ggWWkfactors,HIG16041}.
\DY events are generated with \MGvATNLO v2.2.2~\cite{Alwall:2014hca} at LO, and event yields are scaled to the NNLO cross section~\cite{bib:DYxsec}.
Events from \ttW, \ttZ, triboson, and $\PH\to\WW$ production are generated at NLO~\cite{bib:ttVxsec,bib:HWWxsec} with the \MGvATNLO generator.

Chargino pair production and top squark pair production events are generated using \MGvATNLO at LO with up to two extra partons in the matrix element calculations, and are normalized to the respective cross sections computed at NLO plus next-to-leading logarithmic (NLL) precision~\cite{Beenakker:1999xh,Fuks:2012qx,Fuks:2013vua,Beenakker:1996ch,Kulesza:2008jb,Kulesza:2009kq,Beenakker:2009ha,Beenakker:2011fu,Borschensky:2014cia}, with all the other sparticles assumed to be heavy and decoupled. In the case of chargino pair production, calculations are performed in a limit of mass-degenerate wino \PSGczDt and \PSGcpmDo, and light bino \PSGczDo.

All processes are generated using the NNPDF3.0~\cite{NNPDF} parton distribution function (PDF) set.
The parton showering, hadronization, and the underlying event are modeled using \PYTHIA 8.212~\cite{Sjostrand:2014zea} with the CUETP8M1~\cite{CUETP8M1} underlying event tune for all the processes, except in the generation of \ttbar events, where the first emission is done at the matrix element level with \POWHEG v2 and the CUETP8M2T4\cite{bib:TOP16021} tune is used. Weights for the estimation of theoretical systematic uncertainties, including those related to the choice of PDFs, and renormalization and factorization scales, are included in simulated events~\cite{Kalogeropoulos:2018cke}.

The detector response to the generated events is simulated using a realistic model of the CMS detector based on \GEANTfour~\cite{Agostinelli:2002hh} for SM processes, while for signal events a fast simulation (\FastSim)~\cite{Abdullin:2011zz} of the detector based on a parametrization of the average response to particles is used. Simulated events are subsequently reconstructed using the same algorithms as applied to data.

In order to model the effect of multiple interactions per bunch crossing (pileup), simulated events are mixed with minimum-bias events simulated with \PYTHIA, and are reweighted in order to match the observed rate of multiple interactions.

The modeling and normalization of the main background processes are studied in data, as discussed in Section~\ref{sec:bkgestimation}. The modeling of \ttbar, \tW, and \WW production is studied in data control regions (CRs), and their normalization is determined via a maximum likelihood (ML) fit to data. The normalization of the yields of events from \ttZ, \WZ, \ZZ, and \DY production is taken from simulation and corrected by the event rates measured in dedicated CRs.

To improve the modeling of jets from initial-state radiation (ISR) in simulated signal events, reweighting factors are applied, which make the distributions of observables for related SM processes in simulation agree with control samples in data.
For chargino pair production, mediated by the electroweak interaction, the reweighting procedure is based on studies of \pt balance in inclusive \PZ boson production events~\cite{isrreweightingEKW}. Events are then reweighted according to the total transverse momentum (\ptisr) of the system of supersymmetric particles. The reweighting factors range between 1.18 at $\ptisr\approx 125\GeV$ and 0.78 for $\ptisr>600\GeV$. A global reweighting is further applied in order not to alter the signal production cross section.
As top squark pair production occurs via strong interactions, a different set of reweighting factors is derived as a function of the multiplicity of ISR jets (\Njetisr) in a sample of \ttbar events selected by requiring an OC electron-muon pair and two jets identified as coming from bottom quark hadronization. The measured reweighting factors vary between 0.92 and 0.51 for \Njetisr between 1 and 6, with an additional scale factor applied to keep the total event yields invariant.

\section{Event reconstruction}\label{sec:reconstruction}

The particle-flow algorithm~\cite{CMS-PRF-14-001} aims to reconstruct and identify each individual particle in an event, with an optimized combination of information from the various elements of the CMS detector.
The energy of photons is directly obtained from the ECAL measurement, corrected for zero-suppression effects.
The energy of electrons is determined from a combination of the electron momentum at the primary interaction vertex as determined by the tracker, the energy of the corresponding ECAL cluster, and the energy sum of all bremsstrahlung photons spatially compatible with originating from the electron track. The momentum of muons is obtained from the curvature of the corresponding track. The energy of charged hadrons is determined from a combination of their momentum measured in the tracker and the matching ECAL and HCAL energy deposits, corrected for zero-suppression effects and for the response function of the calorimeters to hadronic showers. Finally, the energy of neutral hadrons is obtained from the corresponding corrected ECAL and HCAL energy.

The reconstructed vertex with the largest value of summed physics-object $\pt^2$ is taken to be the primary \pp interaction vertex. The physics objects are the jets, clustered using a jet finding algorithm~\cite{Cacciari:2008gp,Cacciari:2011ma} with the tracks assigned to the vertex as inputs, and the associated momentum imbalance in the transverse plane, taken as the negative vector \pt sum of those jets.

The identification of the muons used in the analysis is based on the number of reconstructed energy deposits in the tracker and in the muon system, and on the fit quality of the muon track~\cite{Chatrchyan:2012xi}. Electron identification relies on quality criteria of the electron track, matching between the electron trajectory and the associated cluster in the calorimeter, and shape observables of the electromagnetic shower observed in the ECAL~\cite{Khachatryan:2015hwa}. The efficiency for the reconstruction and selection of the muons (electrons) is found to be 70--95 (30--75)\% depending on their \pt and $\eta$.

The lepton selection is further optimized to select leptons from the decays of {\PW} or \PZ bosons. The leptons are required to be isolated by measuring their relative isolation (\Irel), as the ratio of the scalar \pt sum of the photons and of the neutral and charged hadrons within a cone of radius $R = \sqrt{\smash[b]{(\Delta\phi)^2 + (\Delta\eta)^2}}=0.3$ around the candidate lepton, and the \pt of the lepton itself.
The contribution of particles produced in pileup interactions is reduced by considering only charged hadrons consistent with originating from the primary vertex of the event, and correcting for the expected contribution of neutral hadrons from the pileup~\cite{Khachatryan:2015hwa,Chatrchyan:2012xi}. Leptons are considered to be isolated if their relative isolation \Irel is found to be smaller than 0.12. A looser requirement of $\Irel<0.4$ is used to define a veto lepton selection. Candidate lepton trajectories are further required to be compatible with the primary interaction vertex by imposing constraints on their transverse (\dxy) and longitudinal (\dz) impact parameters, and on the three-dimensional impact parameter significance (\Sd), computed as the ratio of the three-dimensional impact parameter and its uncertainty. Both electrons and muons are required to satisfy the conditions $\abs{\dxy}<0.05\cm$, $\abs{\dz}<0.10\cm$, and $\Sd<4$.
Finally, electrons originating from photon conversions are rejected by requiring that the electron track not have missing hits in the innermost layers of the tracker, and  not form a conversion vertex with any other candidate electron in the event~\cite{Khachatryan:2015hwa}.

For each event, hadronic jets are clustered from the PF reconstructed particles using the infrared and collinear-safe anti-\kt algorithm~\cite{Cacciari:2008gp,Cacciari:2011ma} with a distance parameter of 0.4. The jet momentum is determined as the vectorial sum of all particle momenta in the jet, and is found in the simulation to be within 5 to 10\% of its true value over the whole \pt spectrum and detector acceptance.
Jet energy corrections are derived from simulation to bring the measured response of jets to that of particle level jets on average. In situ measurements of the momentum balance in the dijet, multijet, photon+jet, and leptonically decaying {\PZ}+jet events are used to account for any residual difference in jet energy scale in data and simulation~\cite{Khachatryan:2016kdb}.
Additional quality criteria are applied to reject spurious jets from detector noise.
Finally, the jets overlapping with any selected lepton within a cone of radius $R<0.4$ are removed.

Jets originating from the hadronization of bottom quarks (\cPqb\ jets) are identified by the combined secondary vertex v2 \cPqb-tagging algorithm, using the medium operating point~\cite{BTV-16-002}. This requirement provides an efficiency for identifying \cPqb\ jets that increases from 50 to 70\% for jets with \pt from 20 to 100\GeV. The misidentification rate for jets originating from light quarks and gluons is about 1\% in the same \pt range.

The momentum imbalance of the event in the transverse plane is referred to as missing transverse momentum (\ptvecmiss) and it is defined as the negative vectorial \pt sum of all PF candidates in the event, taking into account the energy corrections applied to the jets~\cite{bib:MET}. The magnitude of \ptvecmiss is denoted as \ptmiss.

Differences have been observed in the modeling of the \ptvecmiss resolution in events simulated with \FastSim and with the full detector simulation. To account for this effect, the acceptance for signal events is computed both using the \ptvecmiss at the generator level and after the event reconstruction. The average value of the two acceptances in each analysis bin is taken as the central value for the acceptance.

Simulated events are reweighted to account for differences with respect to data in the efficiencies of the lepton reconstruction, identification, and isolation requirements, and in the performance of \cPqb-jet identification.
The values of the data-to-simulation scale factors differ from unity by less than 10\% with typical efficiency corrections of 2--3 (5)\% for the identification of leptons (\cPqb\ jets) with $\pt>20\GeV$ and $\abs{\eta}<2.4$.

\section{Search strategy}\label{sec:searchstrategy}

The search strategy is developed for two signal hypotheses: the chargino pair and top squark pair productions. The first signal hypothesis is studied along the whole (\invM{\PSGcpmDo}, \invM{\PSGczDo}) mass plane, while for the second one the analysis is optimized on the compressed scenario, where the mass difference of the top squark and the lightest neutralino is in between the top quark and {\PW} boson masses. The searches involve the same techniques for the background estimation and the signal extraction, while they differ slightly in the signal region (SR) selection in order to improve their respective sensitivity.

The signal models are characterized by a common final state with two OC leptons and two lightest neutralinos contributing to large \ptmiss. Based on this, a general high-acceptance baseline selection is defined, requiring two OC isolated leptons with $\abs{\eta}<2.4$ and  $\pt\geq 25$ (20)\GeV for the leading (trailing) lepton.
Events with \PGt leptons decaying into electrons or muons that satisfy the selection requirements are taken into account.
To reduce the contributions from low-mass resonances, $\PZ\to\PGt\PGt$ production, and nonprompt leptons from hadronic jets, the invariant mass \mll of the lepton pair is required to be greater than 20\GeV, and if both leptons have the same flavor (SF), \mll is further required to satisfy $\DZmass>15\GeV$, where \invM{\PZ} is the \PZ boson mass.
High \ptmiss (${\geq}140\GeV$) is required. Events are further rejected if they contain a third lepton with $\pt>15\GeV$, $\abs{\eta}<2.4$, and satisfying the veto lepton selection (as detailed in Section~\ref{sec:reconstruction}). A summary of the baseline selection is found in Table~\ref{Tab:BaseSel}.

\begin{table}[htb]
\centering
\topcaption[center]{Definition of the baseline selection used in the searches for chargino and top squark pair production.}
\begin{tabular}{lcc}
\hline
Variable & \multicolumn{1}{c}{Selection} \\
\hline
 Lepton flavor         & \EE, \MM, \Pepm\PGmmp\\
 Leading lepton        & $\pt\geq 25\GeV$, $\abs{\eta}<2.4$ \\
 Trailing lepton       & $\pt\geq 20\GeV$, $\abs{\eta}<2.4$ \\
 Third lepton veto    &  $\pt\geq 15\GeV$, $\abs{\eta}<2.4$    \\
 \mll                  & ${\geq}20\GeV$ \\
 \DZmass               & ${>}15\GeV$ only for \ee and \mumu events\\
 \ptmiss               & ${\geq}140\GeV$\\
\hline
\end{tabular}
\label{Tab:BaseSel}
\end{table}

The SM processes that contribute most after the baseline selection are \ttbar, \tW, and \WW production. For all these backgrounds, the lepton pair and \ptvecmiss come from a {\PW} boson pair.
Consequently, the variable \MT{T2}~\cite{Lester:1999tx} is constructed to generalize the transverse mass (\MT{T}) for a system with two invisible particles, by using the two leptons as the two visible systems,
\begin{equation}
                                                     \mtll = \min_{\tvec{miss1} + \tvec{miss2} = \ptvecmiss} \left( \max \left[ \MT{T}(\tvec{lep1},\tvec{miss1}), \MT{T}(\tvec{lep2},\tvec{miss2}) \right] \right).
\end{equation}
This observable reaches a kinematic endpoint at the \invM{\PW} for the considered backgrounds. Signal events, instead, present \mtll spectra without such an endpoint because of the additional contribution to the \ptvecmiss given by the neutralinos. The sensitivity of the analysis is further enhanced by dividing the SR in bins of \ptmiss. This allows the analysis not only to exploit the larger tails in the \ptmiss distribution of the signal events, but also to optimize the sensitivity to signals with different mass separation between the produced supersymmetric particle and the LSP. Each \ptmiss bin is in turn divided into events with SF and different flavor (DF) leptons to exploit the smaller contamination from \WZ, \ZZ, and \DY production of the latter.

The SRs are further subdivided based on the specific characteristics of each signal model.
A veto on \cPqb-tagged jets is applied to reject \ttbar, \tW, and \ttZ events in the chargino search.
Selected events in the \ptmiss bins below 300\GeV are then split into two different subregions, depending on the presence of a jet with $\pt>20\GeV$ and $\abs{\eta}<2.4$. This allows for a better discrimination between signal events and top quark background, which still contaminates the SRs after applying the \cPqb-tagged jet veto.
Events with \cPqb-tagged jets are kept as a CR for the normalization of the background from \ttbar and \tW production (discussed in Section~\ref{sec:bkgestimation}).

The final states produced in the top squark decays are characterized by the presence of two bottom quarks.
When the difference in the mass of the top squark and the neutralino is close to the edge of the compressed region, $\Dm\gtrsim\invM{\PW}$, the bottom quarks are soft and give rise to jets with relatively low momentum that have a lower probability to be tagged.
In this case, the top squark final states are similar to those from chargino pair production, and requiring a veto on \cPqb-tagged jets is again an effective strategy to define SRs with reduced contamination from \ttbar, \tW, and \ttZ backgrounds.
For signal scenarios with larger \Dm, instead, the \cPqb\ jets have higher momentum and the final states are more \ttbar-like. Consequently, sensitivity to top squark production is enhanced by requiring a \cPqb-tagged jet to reduce the background from diboson and \DY events.

Another useful means to discriminate top squark production from SM processes is given by the presence of high-\pt jets from ISR in the events. The invisible particles (neutrinos and neutralinos) produced in the decay chain of the top squark in the compressed scenario are expected to be soft; events with harder neutralinos, however, can arise when the top squark pair system recoils against a high-\pt ISR jet. In this hard ISR regime, background is still constrained by the kinematic \invM{\PW} endpoint in \mtll, and can be effectively separated from  the signal.
Hard ISR events are selected by requiring that the leading jet satisfies $\pt>150\GeV$ and is not \cPqb\ tagged. In order to favor the topology in which the jet recoils against the rest of the system, the $\Dphi$ between the jet and the \ptvecmiss is required to be larger than 2.5 rad. This requirement is found to be effective in discriminating top squark production from background events at high \ptmiss, and is therefore applied only for events with $\ptmiss>300\GeV$.

\begin{figure}
\centering
\includegraphics[width=0.48\textwidth]{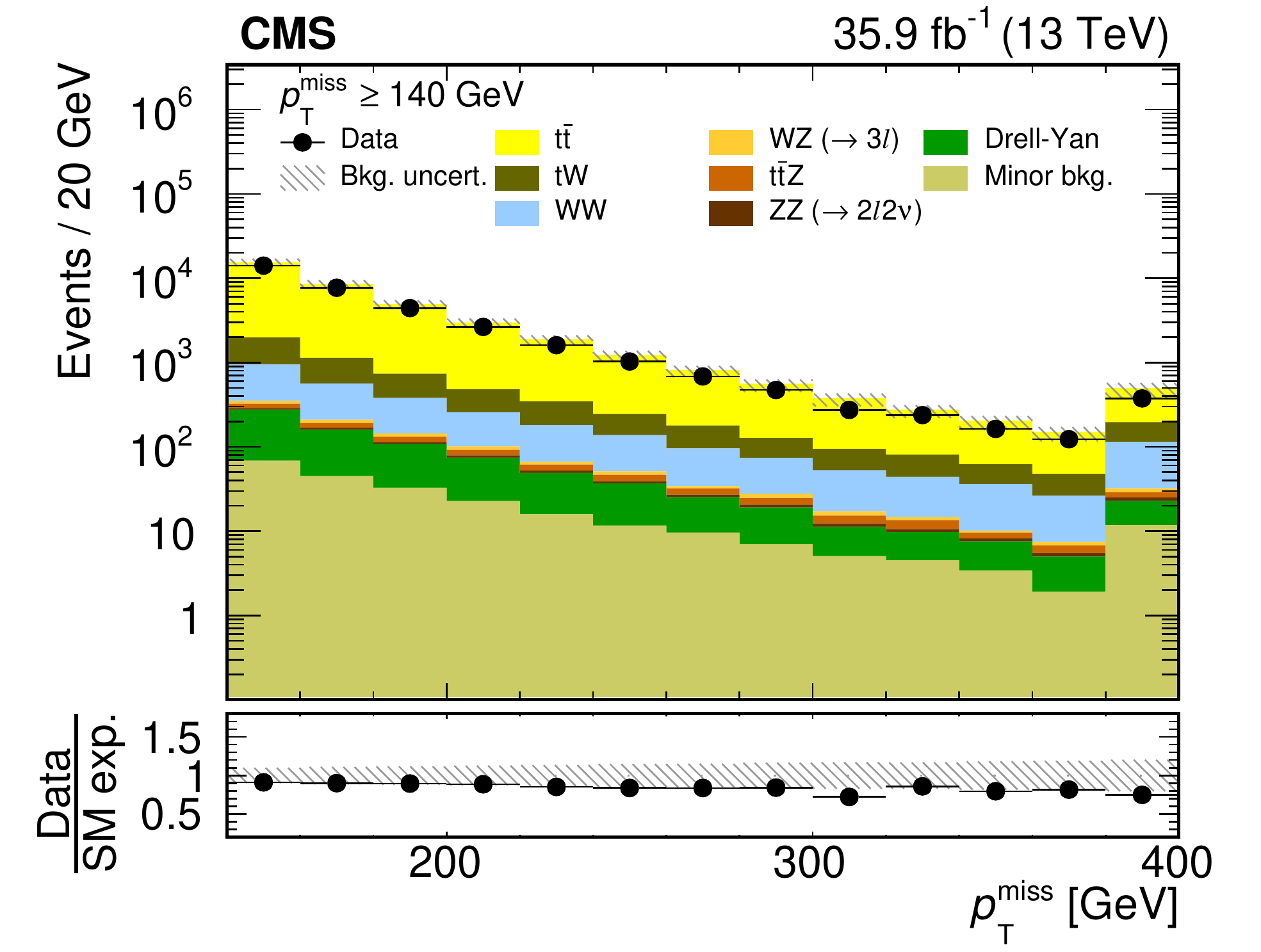}
\includegraphics[width=0.48\textwidth]{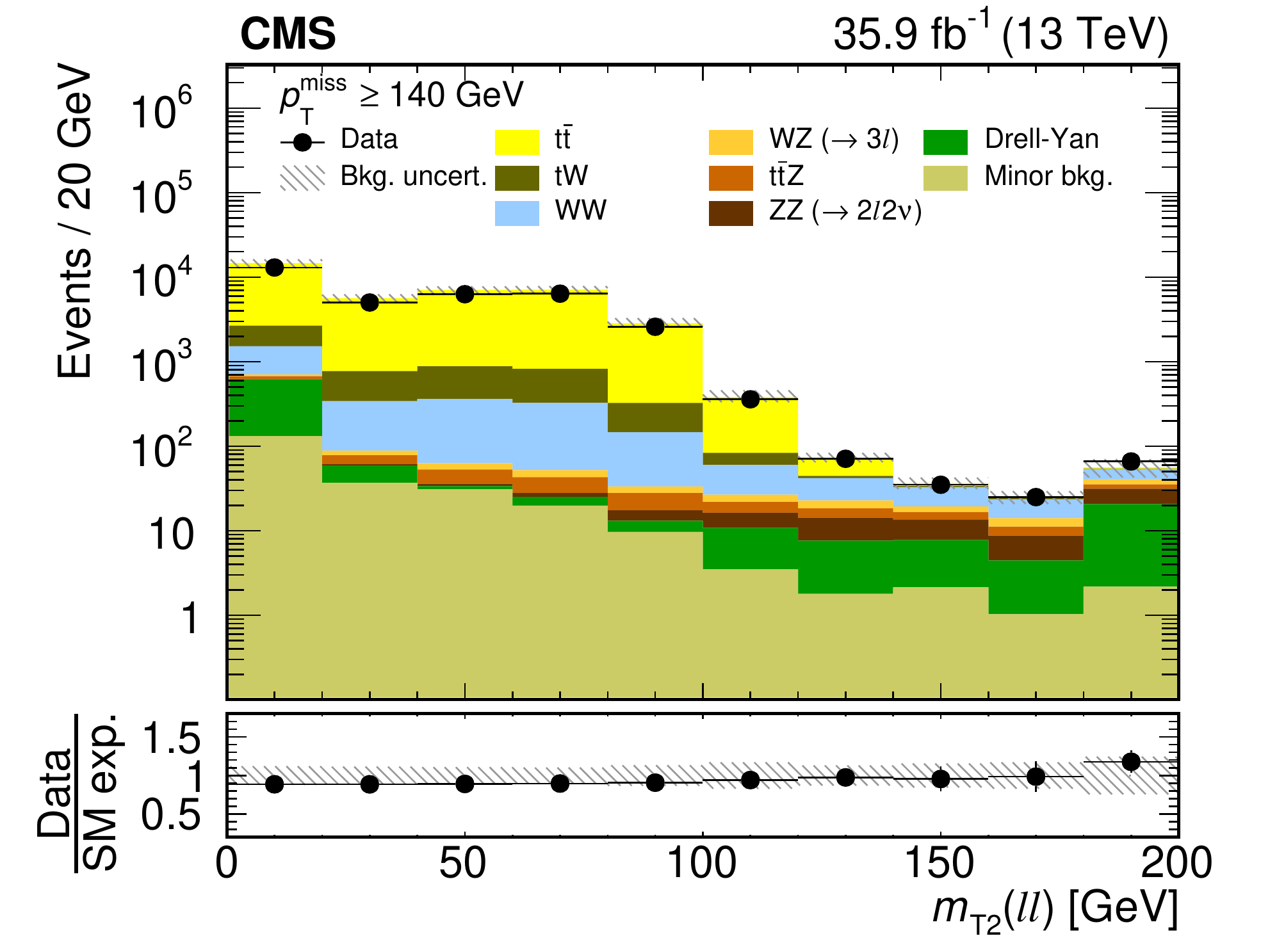}
\includegraphics[width=0.48\textwidth]{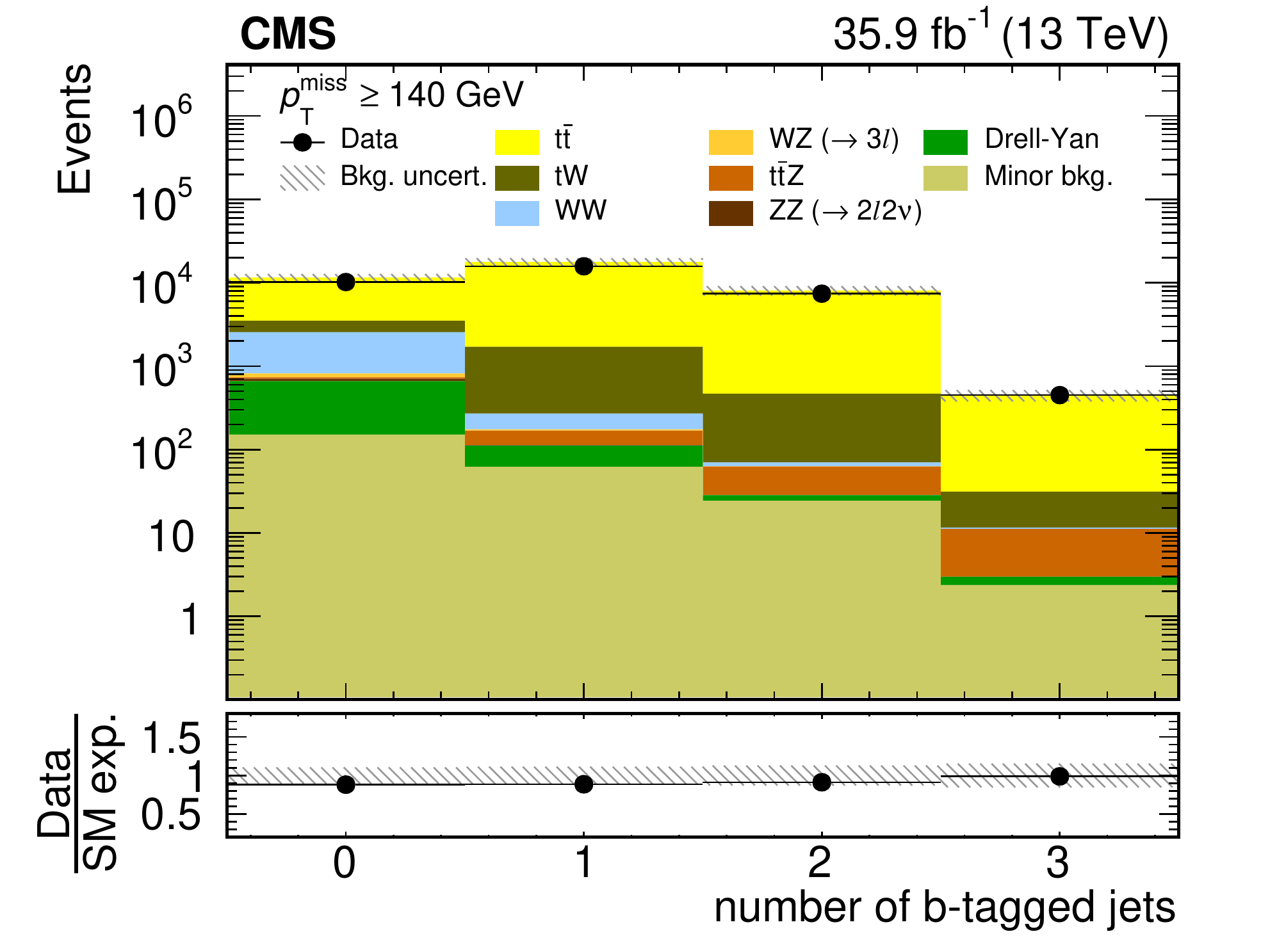}
\includegraphics[width=0.48\textwidth]{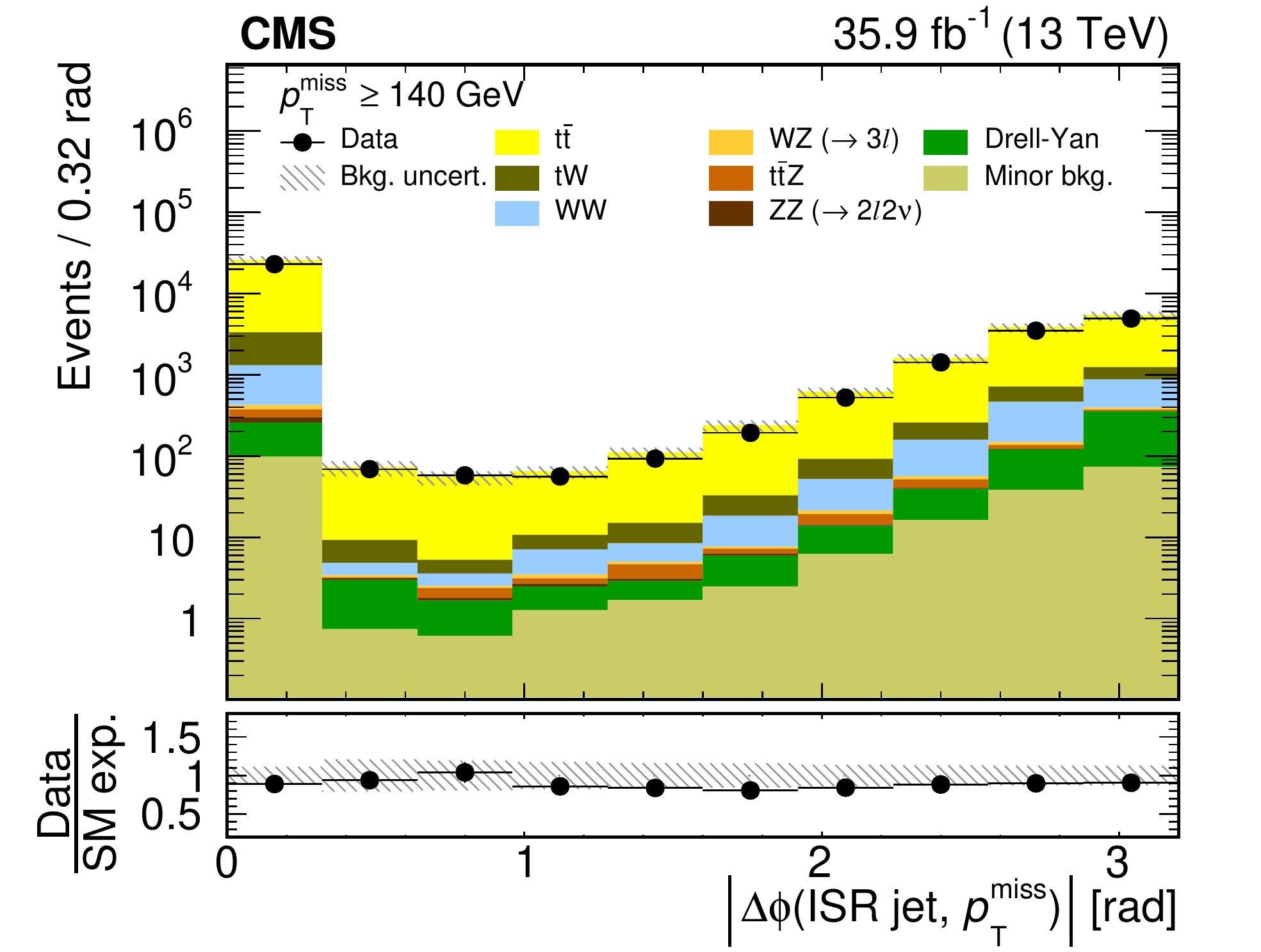}
\caption{Observed and SM expected distributions of some observables used to define the SRs for events with two OC isolated leptons and $\ptmiss\geq 140\GeV$ . Clockwise from top left: \ptmiss, \mtll, \Dphi between the \ptvecmiss and the leading jet (required not to be \cPqb-tagged and with $\pt>150\GeV$, events missing this requirements are shown in the first bin), and multiplicity of \cPqb-tagged jets in the event. The last bin includes the overflow entries. The contributions of minor backgrounds such as \ttW, $\PH\to\WW$, and triboson production are grouped together. In the bottom panel, the ratio of observed and expected yields is shown. The hatched band represents the total uncertainty in the background expectation, as described in Section~\ref{sec:systematics}.}
\label{Fig:SRsKin}
\end{figure}

A summary of the SRs for the chargino and top squark searches is given in Tables~\ref{Tab:TChipmSR} and~\ref{Tab:T2ttSR}, respectively, indicating the \ptmiss range, the selection on the multiplicity of jets (\N{jets}) and \cPqb\ jets (\N{\cPqb\ jets}) in the event, and the ISR jet requirement. The observed distributions of some observables used to define the SRs are compared to SM expectations in Fig.~\ref{Fig:SRsKin}.

\begin{table}[htb]
\centering
  \topcaption[center]{Definition of the SRs for the chargino search as a function of the \ptmiss value, the \cPqb-jet multiplicity and jet multiplicity. Also shown are the CRs with \cPqb-tagged jets used for the normalization of the \ttbar and \tW backgrounds. Each of the regions is further divided in seven \mtll bins as described in the last row.}\label{Tab:TChipmSR}
\cmsTable{
\begin{tabular}{lcccccccc}
\hline
      & \Reg{SR1}{0tag}{0jet} & \Reg{SR1}{0tag}{jets} & \Reg{CR1}{tags}{}  & \Reg{SR2}{0tag}{0jet} & \Reg{SR2}{0tag}{jets} & \Reg{CR2}{tags}{}  & \Reg{SR3}{0tag}{} & \Reg{CR3}{tags}{} \\
\hline
\ptmiss [{\GeVns}]          & 140--200 &  140--200  & 140--200 &  200--300  & 200--300  & 200--300 & ${\geq}300$ &  ${\geq}300$   \\
\N{\cPqb\ jets}  &  0  &  0  & ${\geq}1$ & 0 & 0 & ${\geq}1$  &  0 & ${\geq}1$\\
\N{jets}    &  0  & ${\geq}1$  & ${\geq}1$ & 0 & ${\geq}1$ & ${\geq}1$ & ${\geq}0$ & ${\geq}1$ \\
Channels &  SF, DF   & SF, DF & SF, DF & SF, DF & SF, DF & SF, DF & SF, DF & SF, DF \\ [\cmsTabSkip]
\mtll  & \multicolumn{8}{c}{0--20, 20--40, 40--60, 60--80, 80--100, 100--120, ${\geq}120\GeV$} \\
\hline
\end{tabular}
}
\end{table}

\begin{table}[htb]
 \centering
  \topcaption[center]{Definition of the SRs for top squark production search as a function of the \ptmiss value, the \cPqb-jet multiplicity and the ISR jet requirement. Each of the regions is further divided in seven \mtll bins as described in the last row.}\label{Tab:T2ttSR}
\begin{tabular}{lcccccc}
\hline
      & \Reg{SR1}{0tag}{} & \Reg{SR1}{tags}{} & \Reg{SR2}{0tag}{} & \Reg{SR2}{tags}{} & \Reg{SR3}{0tag}{ISR} & \Reg{SR3}{tag}{ISR}\\
\hline
\ptmiss [{\GeVns}]         & 140--200 &  140--200  &  200--300  & 200--300 &  ${\geq}300$  &  ${\geq}300$  \\
\N{\cPqb\ jets}  & 0 & ${\geq}1$ &0 & ${\geq}1$ & 0 & ${\geq}1$ \\
\N{jets}    &  ${\geq}0$  & ${\geq}1$ & ${\geq}0$ & ${\geq}1$ & ${\geq}1$ & ${\geq}2$ \\
ISR jets  & ${\geq}0$ & ${\geq}0$ & ${\geq}0$ & ${\geq}0$ & ${\geq}1$ & ${\geq}1$ \\
Channels &  SF, DF   & SF, DF & SF, DF & SF, DF & SF, DF & SF, DF  \\ [\cmsTabSkip]
\mtll  & \multicolumn{6}{c}{0--20, 20--40, 40--60, 60--80, 80--100, 100--120, ${\geq}120\GeV$} \\
\hline
\end{tabular}
\end{table}

Each of the SRs defined in Tables~\ref{Tab:TChipmSR} and~\ref{Tab:T2ttSR} is further divided into seven \mtll bins of 20\GeV width, starting from 0\GeV and with the last bin collecting all events with $\mtll>120$\GeV. A simultaneous ML fit to the \mtll distribution in all the SRs is then performed to extract the signal (as described in Section~\ref{sec:results}). Since the first \mtll bins have a low signal contribution, we exploit them to constrain the contributions of the dominant backgrounds in the SRs with one \cPqb-tagged jet (dominated by \ttbar and \tW production) and without \cPqb-tagged jets (where \WW production becomes relevant) through the fit.

\section{Background estimation}\label{sec:bkgestimation}

The main contributions from SM processes to the SRs comes from \ttbar, \tW, and \WW production. The normalization of these backgrounds is determined by the ML fit, as mentioned in Section~\ref{sec:searchstrategy}. Their \mtll shape has a natural endpoint at the \invM{\PW},
and events enter into the relevant region for signal extraction ($\mtll>80\GeV$) mainly due to detector resolution effects, whose contributions are not easy to model.
For this reason, we study the modeling of the \mtll distribution for these processes in dedicated CRs in data described in Section~\ref{sec:mt2llShape}. The contributions of  the subleading \ttZ, \WZ, \ZZ, and \DY backgrounds are also tested in CRs, where correction factors for their normalizations are extracted, as discussed in Section~\ref{Sec:bckNorma}.
Remaining minor backgrounds from \ttW, $\PH\to\WW$, and triboson production give small contributions in the SRs, and the estimates for these processes are taken directly from simulation.
Background contributions from rest of the SM processes are found to be negligible.
The contribution of signal to any of the CRs used is found to be negligible compared to SM processes.

\subsection{Modeling of \texorpdfstring{\mtll}{MT2(ll)} in \texorpdfstring{\ttbar}{ttbar}, \tW, and \WW events}\label{sec:mt2llShape}

The simulated \mtll distributions for \ttbar, \tW, and \WW backgrounds are validated in two CRs.
To construct the first one, the baseline selection is modified by requiring  $100<\ptmiss<140\GeV$.
The events in this CR are further separated according to their \cPqb-jet multiplicity to define two sub-regions with different content in top quark (\ttbar and \tW) and \WW backgrounds.
In order to reject events from \DY production, only DF events are considered.
The second CR aims at validating the modeling of the \mtll distributions in events with $\ptmiss>140\GeV$. For this purpose,
we select events from $\WZ\to 3\ell1\nu$ production and emulate the \mtll shape of \WW and top quark events.  We take the lepton from the \PZ boson with the same charge as the lepton from the {\PW} boson, and we add its \pt vectorially to \ptvecmiss, effectively treating it like a neutrino.
 These events are selected by requiring three leptons and vetoing the presence of a fourth lepton passing the veto lepton requirements. A veto is applied to events with \cPqb-tagged jets to remove residual \ttbar events. Among the three leptons, a pair of OC SF leptons with an invariant mass within 10\GeV of the \PZ boson mass is required to identify the \PZ boson.
The simulation is found to describe the data well in the CRs.
Based on the statistical precision of these CRs, a conservative uncertainty of 5, 10, 20, and 30\% is taken for the bins $60\leq\mtll <80\GeV$, $80\leq\mtll <100\GeV$, $100\leq\mtll <120\GeV$, and $\mtll\geq 120\GeV$, respectively. These uncertainties are applied to top quark and \WW production, and treated as uncorrelated between the two types of backgrounds.

Another potential source of mismodeling in the tails of the \mtll distributions arises from nonprompt leptons originating, for instance, from semileptonic decays of {\PB} hadrons in \cPqb\ jets or from hadronic jets accidentally passing the lepton selection. The value of \mtll in \ttbar, \tW, and \WW events with one nonprompt lepton replacing a prompt one failing the selection requirements will not be bound by the \invM{\PW} endpoint.
The contribution of these events is found to be less than 1\% of the expected background across the different SRs. It becomes more relevant only at large values of \mtll and \ptmiss, where it constitutes up to 20\% of the \ttbar background.
We study the modeling of the rate of nonprompt leptons in simulation by selecting events with two leptons with the same charge and at least one \cPqb-tagged jet.
The dominant contribution to this sample comes from \ttbar events with a nonprompt lepton.
Based on the observed agreement with data, a correction factor of $1.08\pm 0.21$ is derived for the nonprompt lepton rate in simulation.

\subsection{Normalization of \texorpdfstring{\ttZ}{ttZ}, \WZ, \ZZ, and \DY backgrounds}\label{Sec:bckNorma}

The production of \ttZ events where the two {\PW} bosons decay leptonically and the \PZ boson decays into neutrinos leads to final states with the same experimental signature as the signal events and with no natural endpoint for the reconstructed \mtll distribution, due to the additional contribution of the neutrinos from the \PZ boson decay to the \ptvecmiss. The normalization of this background is validated in events with three leptons, $\ptmiss> 140\GeV$, and at least two jets with $\pt> 20\GeV$, of which at least one is tagged as \cPqb\ jet. At least one pair of OC SF leptons with an invariant mass not further than 10\GeV from the \PZ boson mass is also required.  A normalization scale factor of $1.44\pm 0.36$ for \ttZ production is measured comparing the observed and predicted numbers of events.

Events from \WZ production enter the signal event selection when both bosons decay leptonically and one of the three decay leptons fails the veto lepton requirements. We test the modeling of this source of background in a CR with three leptons, $\ptmiss>140\GeV$, and no \cPqb-tagged jets, and derive a normalization scale factor of $0.97\pm 0.09$ for the simulated \WZ background.

The \ZZ background is dominated by events with one boson decaying into charged leptons and the other one decaying into neutrinos. This contribution is studied by mimicking the $\ZZ\to 2\ell2\nu$ production via $\ZZ\to 4\ell$  events, where the \pt of one of the reconstructed \PZ bosons (randomly chosen between the ones satisfying the $\DZmass<15$\GeV condition) is added to the \ptvecmiss. Events are selected by requiring four leptons, with one lepton allowed to pass the looser veto lepton requirement in order to increase the acceptance for \ZZ production. The events are retained if the four leptons can be arranged into two pairs of OC SF leptons, both with an invariant mass within 30\GeV of the \PZ boson mass, and at least one within 15\GeV. A scale factor for the \ZZ background normalization is derived in events with $\ptmiss>140\GeV$ and with no \cPqb-tagged jets. Since the chargino search uses separate SRs for events with or without jets, two corresponding scale factors are also measured,
which suggest a higher jet multiplicity in data than in \ZZ simulated events.

A summary of the scale factors derived in this section is given in Table~\ref{Tab:SFofNormaliza}. For all the quoted scale factors, uncertainties include the statistical uncertainties on data and simulated events, and the systematic uncertainties on the number of expected events from the residual processes in the CRs.

\DY events can pass the baseline selection because of mismeasurements in \ptmiss. We study the modeling of this background in events with two OC SF leptons with $\DZmass<15$\GeV, no additional leptons, and no \cPqb-tagged jets (Z boson events). The events with $100<\ptmiss<140$\GeV are dominated by \DY production, and are used to derive a \mtll shape correction, which is subsequently tested in \PZ boson events with $\ptmiss>140$\GeV. The correction ranges from a few percent at low \mtll to about 50\% for $\mtll>100\GeV$. An overall  normalization uncertainty of 32\% is also established by the observed disagreement between data and simulated events with $100<\ptmiss<140$\GeV.
Finally, the predictions for \DY events with no jets are tested in \PZ boson events with no jets and $\ptmiss>140\GeV$: a conservative uncertainty of 100\% in this contribution is applied.
The \DY production is a subdominant background in the SRs with no jets and this uncertainty has a negligible impact on the expected sensitivity for signal production.

\begin{table}[htb]
 \centering
  \topcaption[center]{Summary of the normalization scale factors for \ttZ, \WZ, and \ZZ backgrounds in the SRs used for the chargino (a) and top squark (b) searches. Uncertainties include the statistical uncertainties of data and simulated event samples, and the systematic uncertainties on the number of expected events from the residual processes in the CRs.}
  \label{Tab:SFofNormaliza}
\begin{tabular}{lccc}
\hline
\multirow{2}{*}{Process} & \multicolumn{3}{c}{Scale factors} \\
  &  $\N{jets} = 0$ (a) & $\N{jets} > 0$ (a)  &  $\N{jets} \geq 0$ (b) \\
\hline
\ttZ     & $1.44 \pm 0.36$ & $1.44 \pm 0.36$  & $1.44 \pm 0.36$ \\
\WZ      & $0.97 \pm 0.09$ & $0.97 \pm 0.09$  & $0.97 \pm 0.09$ \\
\ZZ     & $0.74 \pm 0.19$ & $1.21 \pm 0.17$ & $1.05 \pm 0.12$  \\
\hline
\end{tabular}
\end{table}

\section{Systematic uncertainties}\label{sec:systematics}

Several sources of systematic uncertainty that affect both the normalizations and the \mtll shapes of the background and signal events are considered in the analysis.
\begin{itemize}
\item The overall uncertainty in the integrated luminosity is estimated to be 2.5\%~\cite{CMS-PAS-LUMI-2017}.
\item The uncertainty on the measured trigger efficiency is 2\%.
\item Lepton identification and isolation efficiencies are corrected by data-to-simulation scale factors measured in $\PZ\to\ell\ell$ events. The corresponding uncertainties are typically smaller than 3\% per lepton.
\item The jet energy scale is varied by its uncertainty~\cite{Khachatryan:2016kdb}, and the changes are propagated to all the related observables in the event.
\item The energy scale of the low-\pt particles that are not clustered in jets is varied by its uncertainty, and the changes are propagated to the \ptvecmiss.
\item The efficiencies and misidentification rates of the \cPqb-jet identification algorithms are also corrected by data-to-simulation scale factors measured in inclusive jet and \ttbar events~\cite{BTV-16-002}. The respective uncertainties range between 1 and 6\%, depending on the \pt and $\eta$ of the jets.
\item The effect of the simulated data sample sizes on the modeling of the \mtll distributions is taken into account by treating the statistical uncertainty in each bin for each process as an additional uncorrelated uncertainty.
\item Uncertainties in the renormalization and factorization scales, and PDFs are propagated by taking the largest changes in the acceptance when independently doubling and halving the renormalization and factorization scales, and when varying the choice of PDFs between the NNPDF3.0 replicas. The PDF uncertainties are not considered for signal models as they are found to be redundant, once the uncertainty in the ISR modeling is included.
\end{itemize}
The estimates of the SM backgrounds are also affected by specific uncertainties in the modeling of the different processes.
\begin{itemize}
\item A background normalization uncertainty is applied for each background separately. The normalizations of the \ttbar, \tW, and \WW processes are determined by the ML fit, as described in Section~\ref{sec:results}. We assign a common normalization parameter for \ttbar and  \tW events and another for \WW production. No explicit normalization uncertainty is defined for \ttbar and \WW events, while a 10\% uncertainty is set for the \tW process to take into account its relative normalization with respect to the \ttbar production as well as any interference effect between them. The uncertainties applied to \ttZ (25\%), \WZ (9\%), and \ZZ (26\% in the SRs with 0 jets, 14\% in the SRs with at least 1 jet, and 11\% in the rest of the SRs) correspond to the scale factor uncertainties obtained in Section~\ref{Sec:bckNorma}. Minor backgrounds (\ttW, $\PH\to\WW$, triboson production) are assigned a conservative uncertainty of 50\%. Finally, \DY events have a 100\% normalization uncertainty in the SR with no jets and 32\% in all other SRs.
\item The modeling of the yields of events with no jets has been explicitly studied in Section~\ref{Sec:bckNorma} for \ZZ and \DY production. For the other SM processes, we introduce a related uncertainty by adding two free parameters in the ML fit, scaling respectively the rate of events with no jets for diboson and b-enriched (\ttbar, \tW, \ttZ, and \ttW) backgrounds. The total number of expected events without \cPqb-tagged jets is constrained to remain invariant, so that only a migration of events between the SRs with and without jets is allowed.
\item The modeling of the \mtll  shapes in events with an endpoint at the \invM{\PW} (\ttbar, \tW, and \WW) has been studied in Section~\ref{sec:mt2llShape}: an uncertainty of 5, 10, 20, and 30\% is assigned for the last four \mtll bins.
\item The choice of the set of NNLO/NLO \emph{K} factors applied to the $\Pq\Paq\to\PZ\PZ$ events affects the modeling of the \mtll shape for the \ZZ background (as described in Section~\ref{sec:datasets}). Relative variations range from 16\% for $\mtll<20\GeV$ to about 2\% for $\mtll>120\GeV$ and are taken as the uncertainties.
\item The \mtll distribution in \DY events has been corrected by scale factors derived in bins of \mtll in the validation region $100<\ptmiss<140$\GeV, as discussed in Section~\ref{Sec:bckNorma}. The full size of the correction in each bin is taken as an uncertainty.
\item The weight of events with nonprompt leptons in simulated samples is varied by the ${\pm}19\%$ uncertainty in the correction factor derived in events with two same-charge leptons, as described in Section~\ref{sec:mt2llShape}.
\item The spectrum of top quark \pt in \ttbar events has been observed to be softer in data than in simulated events~\cite{bib:toppt1,bib:toppt3,bib:toppt4}. An uncertainty is derived from the observed variations when reweighting the \ttbar events to the \pt distribution observed in data.
\end{itemize}
Finally, additional uncertainties in the modeling of signal events are taken into account, mostly related to the performance of the event reconstruction in \FastSim.
\begin{itemize}
\item The uncertainty in the lepton identification efficiency in events simulated with \FastSim, relative to the full detector simulation, is estimated to be 2\%.
\item The analogous uncertainty in the \cPqb-tagging efficiency in \FastSim samples ranges between 0.2--0.5\%.
\item The \ptvecmiss modeling in \FastSim events is studied by comparing the acceptances computed using the \ptvecmiss at the generator level and after the event reconstruction. Since the average of the two is taken as central value for the acceptance, half of their difference is taken as an uncertainty, fully correlated among bins.
\item An uncertainty in the modeling of pileup events in \FastSim signal samples is derived by studying the dependence of the acceptance on the multiplicity of primary vertices reconstructed in the event. This uncertainty varies from 0 to 10\% across the SRs and \mtll bins.
\item Simulated signal events are reweighted to improve the modeling of the ISR, as described in Section~\ref{sec:datasets}. Uncertainties on the reweighting procedure are derived from closure tests. For chargino models, the deviation from unity is taken as the systematic uncertainty in the \ptisr reweighting factors. For top squark models, half of the deviation from unity in the \Njetisr factors is taken.
\end{itemize}
Tables~\ref{tab:BackSyst} and~\ref{tab:SignSyst} summarize the systematic uncertainties in the predicted yields for SM processes and for two reference signals, respectively.

\begin{table}[htb]
\centering
\topcaption{Sizes of systematic uncertainties in the predicted yields for SM processes. The first column shows the range of the uncertainties in the global background normalization across the different SRs. The second column quantifies the effect on the \mtll shape. This is computed by taking the maximum variation across the \mtll bins (after renormalizing for the global change of all the distribution) in each SR. The range of this variation across the SRs is given.}
\begin{tabular}{lcc}
\hline
 \multirow{2}{*}{Source of uncertainty}   & \multicolumn{2}{c}{SM processes} \\
  & Change in yields & Change in \mtll shape \\
\hline
Integrated luminosity & 2.5\% & \NA \\
Trigger & 2\% & \NA \\
Lepton ident./isolation & 4--5\%    & ${<}1\%$     \\
Jet energy scale & 1--6\%    & 3--15\%     \\
Unclustered energy & 1--2\%    & 2--16\%     \\
{\cPqb} tagging & ${<}3\%$    & ${<}2\%$     \\
Renorm./fact. scales & 1--10\%    & 1--6\%     \\
PDFs & 1--5\%    & 2--8\%     \\
\ttZ normalization & ${<}1\%$    & ${<}9\%$     \\
\WZ normalization & ${<}1\%$    & ${<}1\%$     \\
\ZZ normalization & ${<}1\%$    & ${<}5\%$     \\
\DY normalization & ${<}4\%$    & 1--11\%     \\
\mtll shape (top quark) & \NA & 4--18\%     \\
\mtll shape (\WW) & \NA & 1--15\%     \\
\ZZ \emph{K} factors & \NA    & ${<}3\%$     \\
\mtll shape (\DY) & \NA    & 1--13\%     \\
Nonprompt leptons & ${<}1\%$    & ${<}4\%$     \\
\ttbar \pt reweighting & 1--4\%    & 1--8\%     \\
 \hline
\end{tabular}
\label{tab:BackSyst}
\end{table}

\begin{table}[htb]
\centering
\topcaption{Same as in Table~\ref{tab:BackSyst} for two representative signal points, one for chargino pair production and one for top squark pair production. }
\cmsTable{
\begin{tabular}{lcc@{\extracolsep{4pt}}cc}
\hline
 \multirow{3}{*}{Source of uncertainty}   & \multicolumn{2}{c}{\TChipmDecay} & \multicolumn{2}{c}{$\stone\to \cPqt\PSGczDo$} \\
    & \multicolumn{2}{c}{($\invM{\PSGcpmDo}=500\GeV$, $\invM{\PSGczDo}=200\GeV$)} & \multicolumn{2}{c}{ ($\invM{\stone}=350\GeV$, $\invM{\PSGczDo}=225\GeV$)} \\
  & \,\,\,\,\,\,Yields\,\,\,\,\,\,\, & \mtll shape & \,\,\,\,\,\,Yields\,\,\,\,\,\,\, & \mtll shape \\
\hline
Integrated luminosity & 2.5\% & \NA  & 2.5\% & \NA \\
Trigger & 2\% & \NA  & 2\% & \NA \\
Lepton ident./isolation  & 4--5\%    & ${<}1\%$  & 4--5\%    & ${<}1\%$    \\
Jet energy scale & 1--3\%    & 3--11\%   & 1--4\%    & 2--14\%    \\
Unclustered energy & 1--2\%    & 8--13\%  & 1--2\%    & 2--7\%      \\
{\cPqb} tagging & ${<}1\%$    & ${<}1\%$  & $1$--$3\%$    & ${<}1\%$    \\
Renorm./fact. scales  & 1--3\%    & 1--3\%  & 1--3\%    & 1--3\%    \\
Lept. id./iso. (\FastSim) & 4\% & ${<}1\%$ & 4\% & ${<}1\%$ \\
{\cPqb} tagging  (\FastSim) & ${<}1\%$ & ${<}1\%$ & ${<}1\%$ & ${<}1\%$ \\
\ptvecmiss (\FastSim)  & 1--4\% & 7--28\% & 1--6\% & 6--20\% \\
Pileup (\FastSim)    & 1--6\% & 4--9\% & 2--4\% & 2--14\% \\
ISR reweighting  & 1--2\% & 1--6\% & 2--8\% & 1--6\% \\
 \hline
\end{tabular}
}
\label{tab:SignSyst}
\end{table}

\section{Results and interpretation}\label{sec:results}

\begin{figure}
\centering
\includegraphics[width=0.48\textwidth]{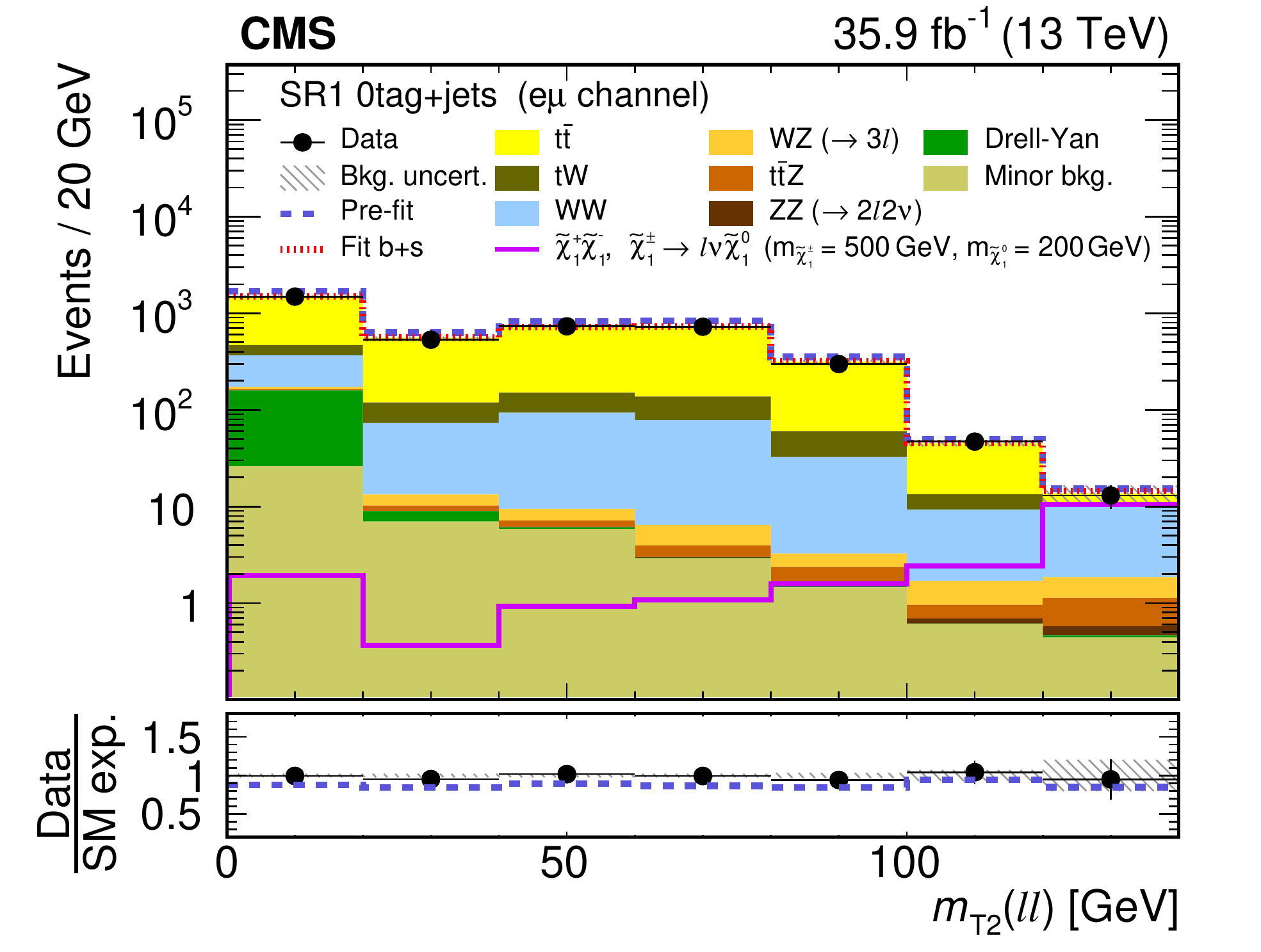}
\includegraphics[width=0.48\textwidth]{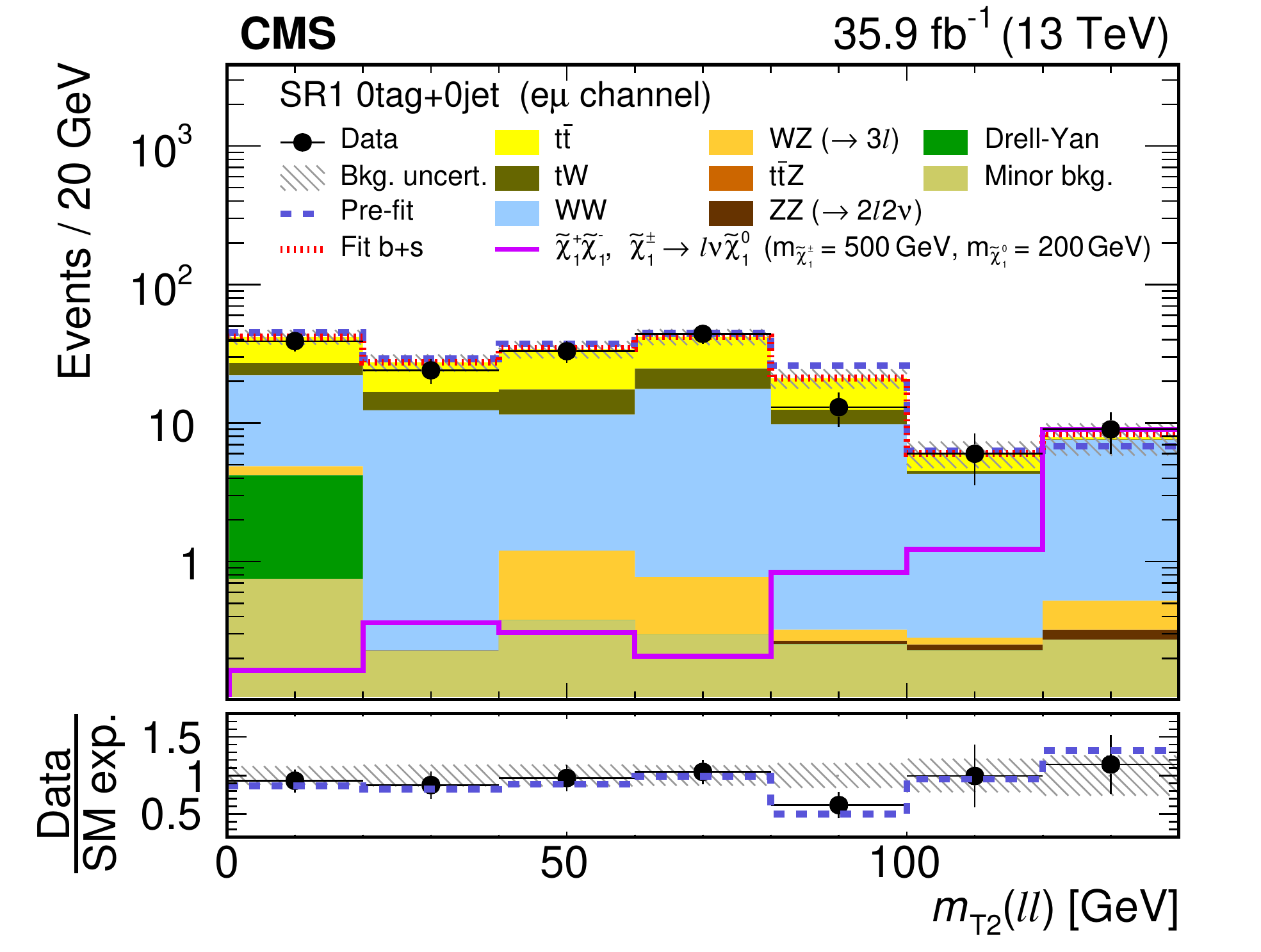}
\includegraphics[width=0.48\textwidth]{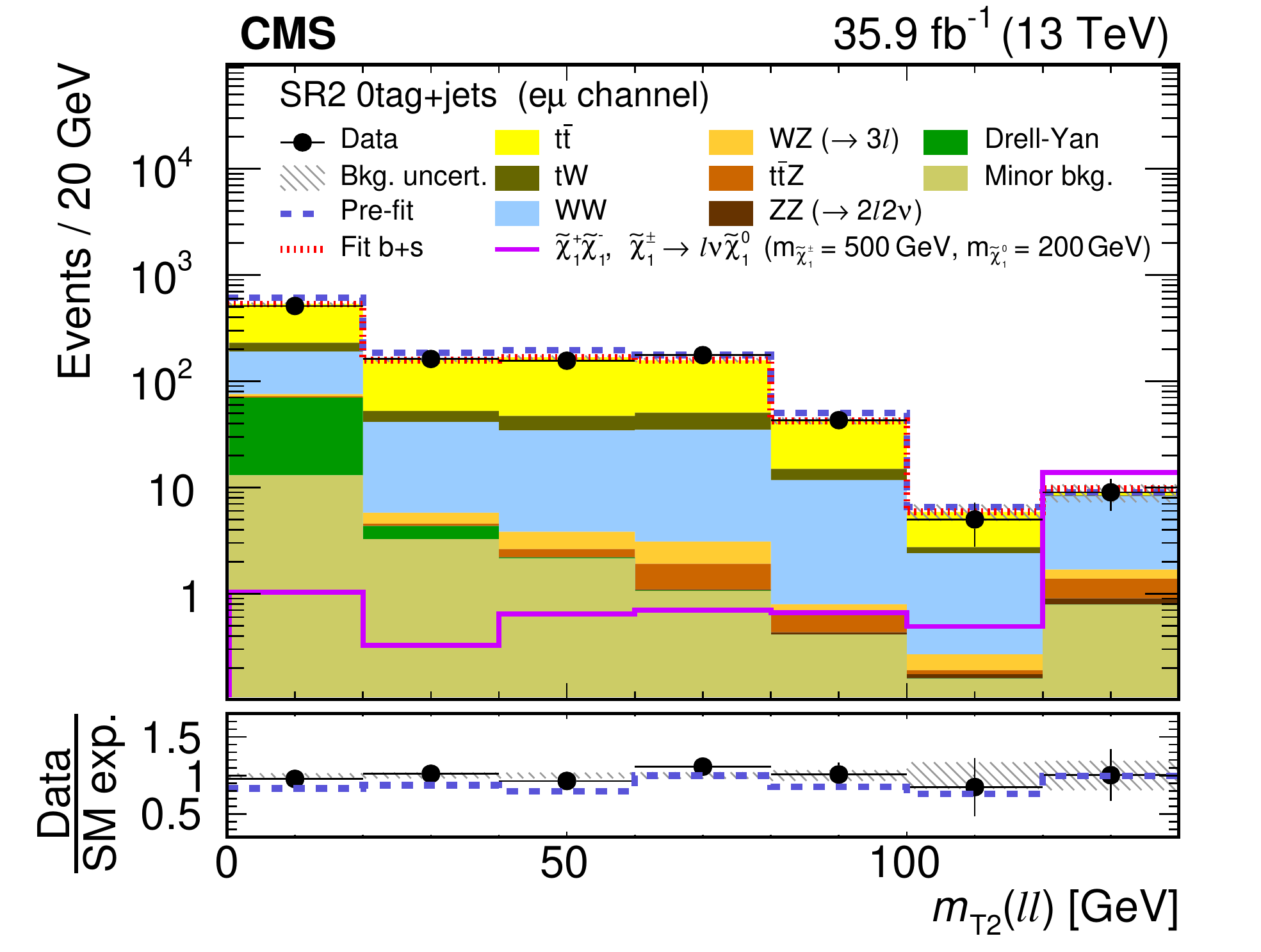}
\includegraphics[width=0.48\textwidth]{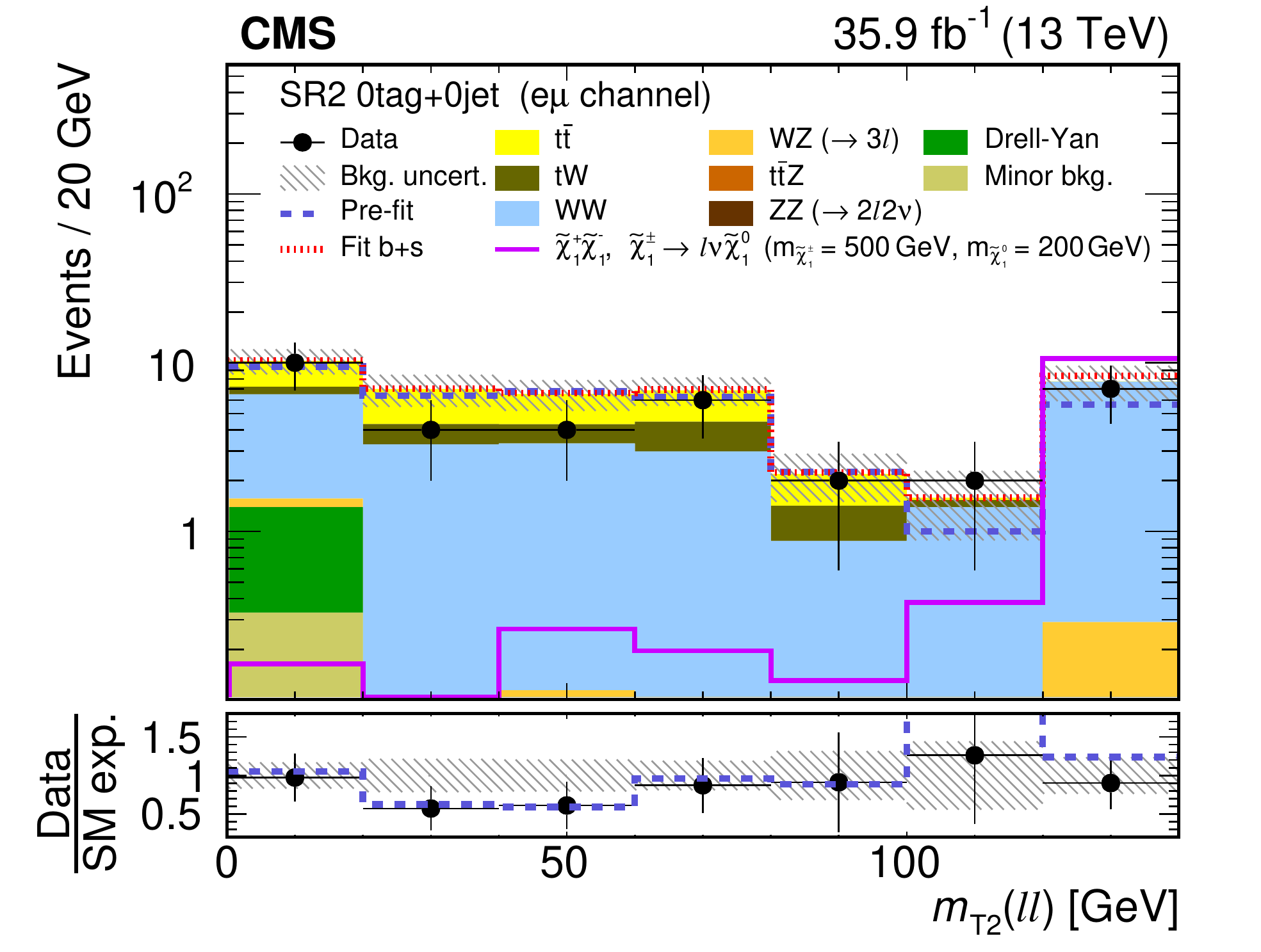}
\includegraphics[width=0.48\textwidth]{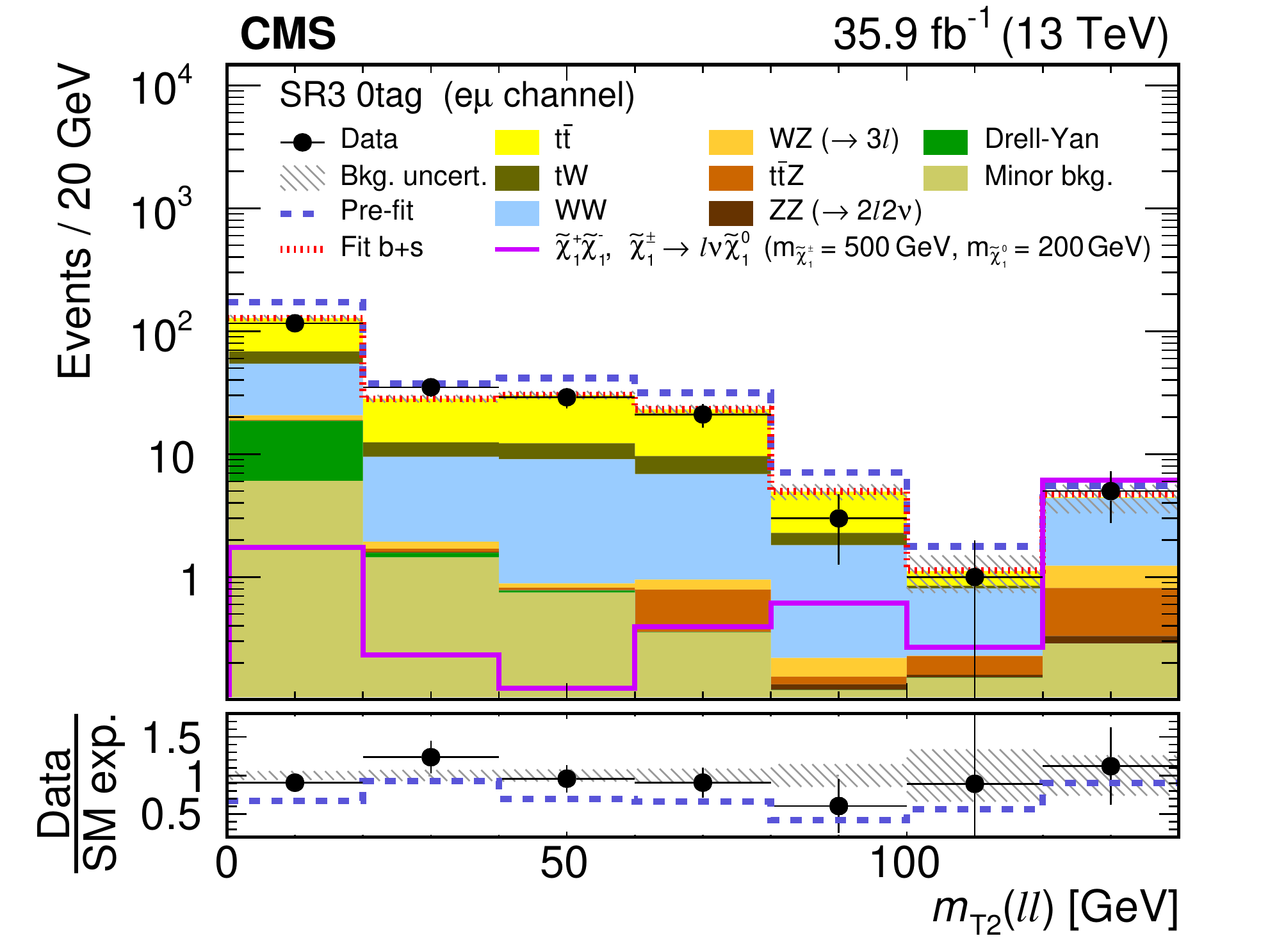}
\caption{Distributions of \mtll after the fit to data in the chargino SRs with $140<\ptmiss<200\GeV$ (upper plots), $200<\ptmiss<300\GeV$ (middle), and $\ptmiss>300\GeV$ (lower), for DF events without \cPqb-tagged jets and at least one jet (left plots) and no jets (right plots). The lower plot for the SR with $\ptmiss>300\GeV$ shows all the events without \cPqb-tagged jets regardless of their jet multiplicity. The solid magenta histogram shows the expected \mtll distribution for chargino pair production with $\invM{\PSGcpmDo}=500\GeV$ and $\invM{\PSGczDo}=200\GeV$. Expected total SM contributions before the fit (dark blue dashed line) and after a background+signal fit (dark red dotted line) are also shown. The last bin includes the overflow entries. In the bottom panel, the ratio of data and SM expectations is shown for the expected total SM contribution after the fit using the background-only hypothesis (black dots) and before any fit (dark blue dashed line). The hatched band represents the total uncertainty after the fit.}
\label{Fig:MT2ll_TChi_em}
\end{figure}

\begin{figure}
\centering
\includegraphics[width=0.48\textwidth]{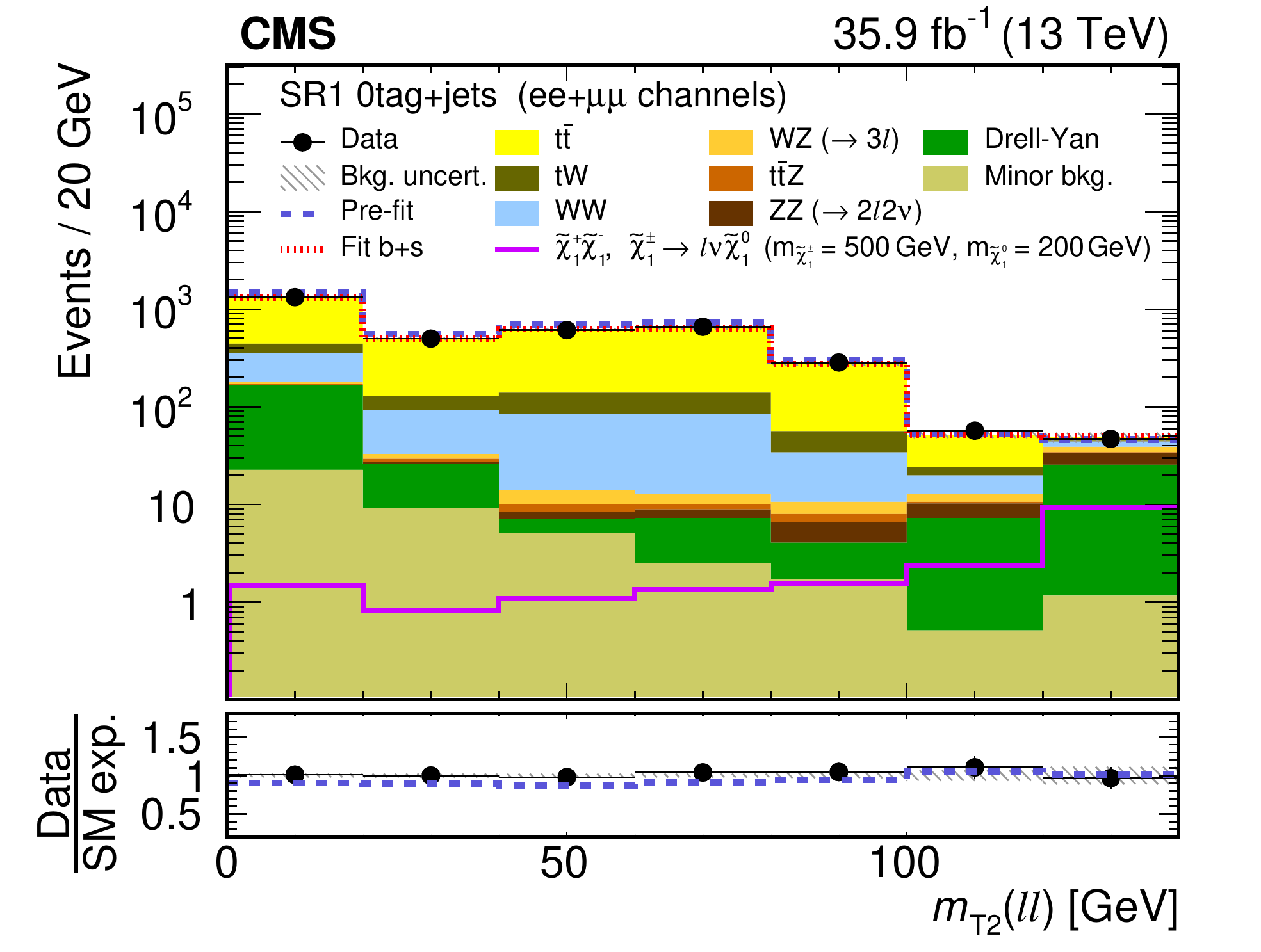}
\includegraphics[width=0.48\textwidth]{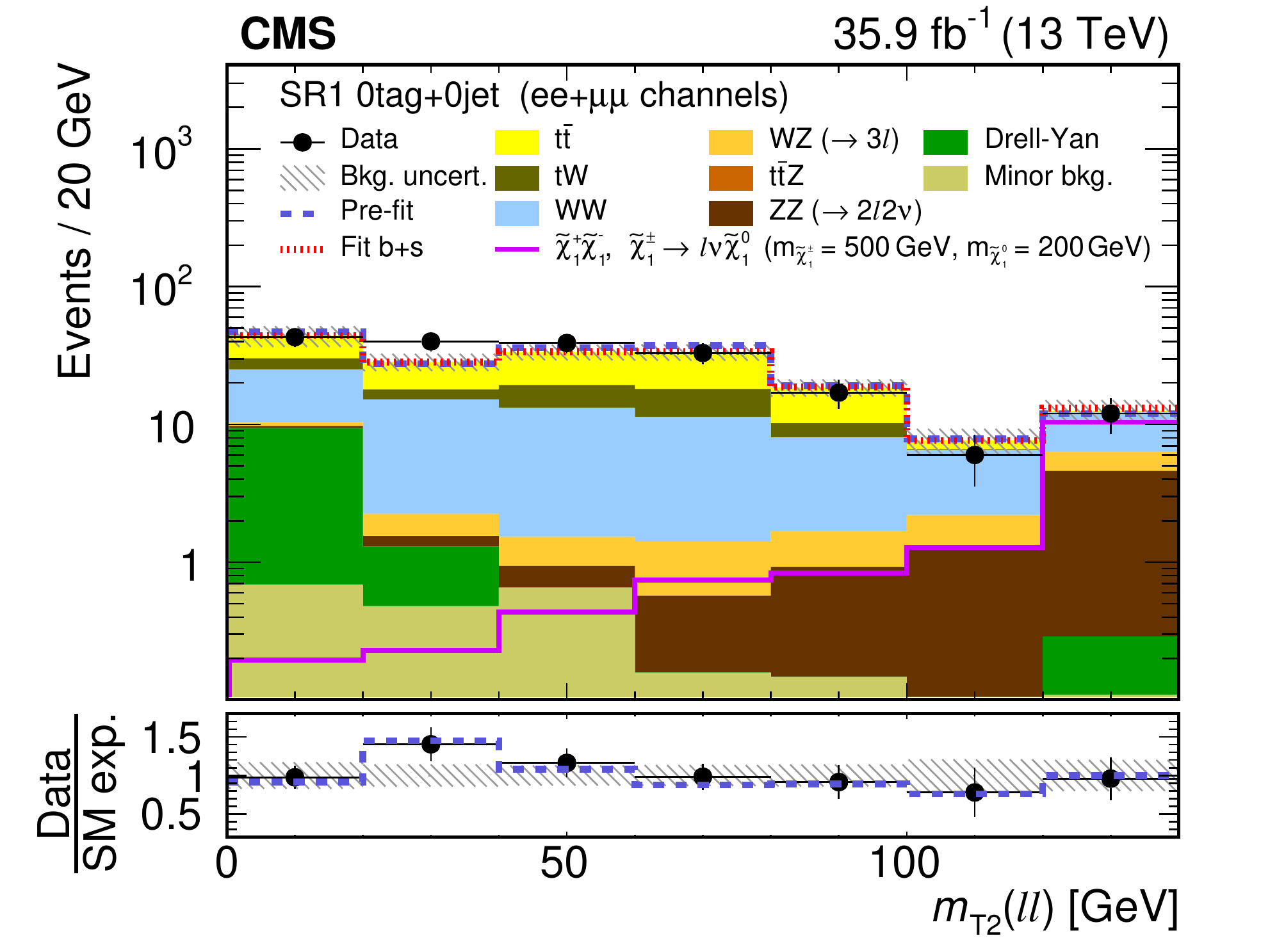}
\includegraphics[width=0.48\textwidth]{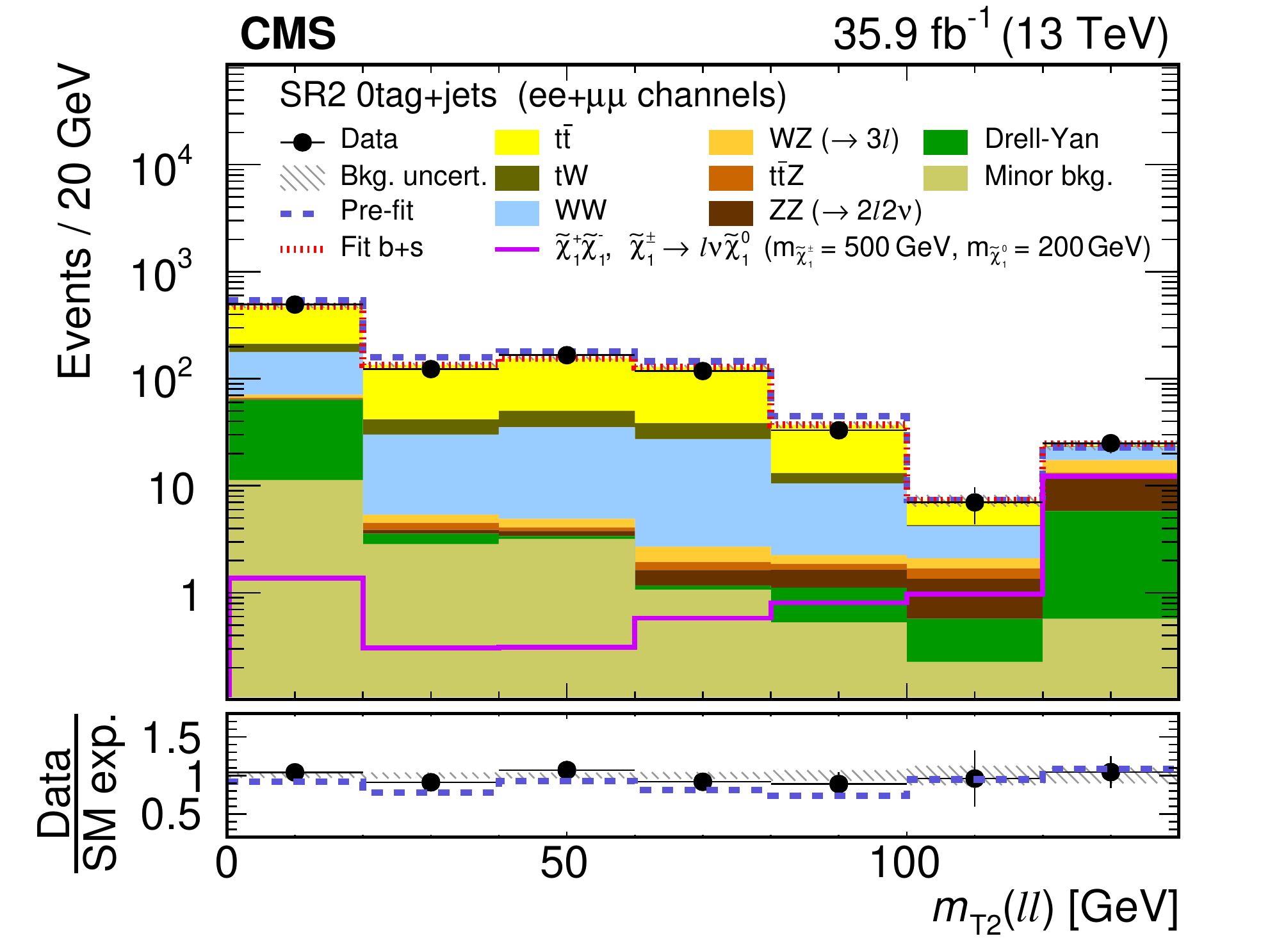}
\includegraphics[width=0.48\textwidth]{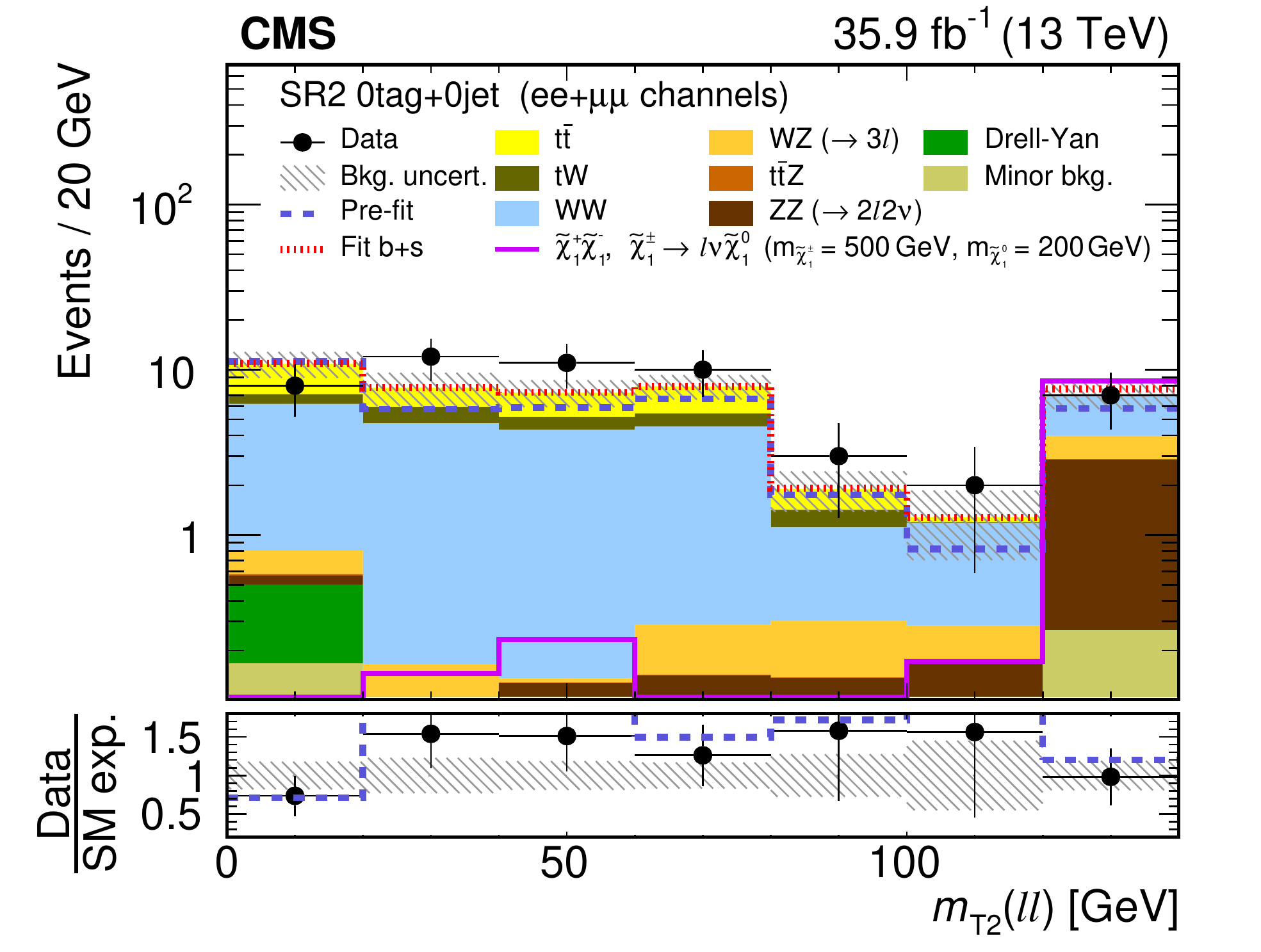}
\includegraphics[width=0.48\textwidth]{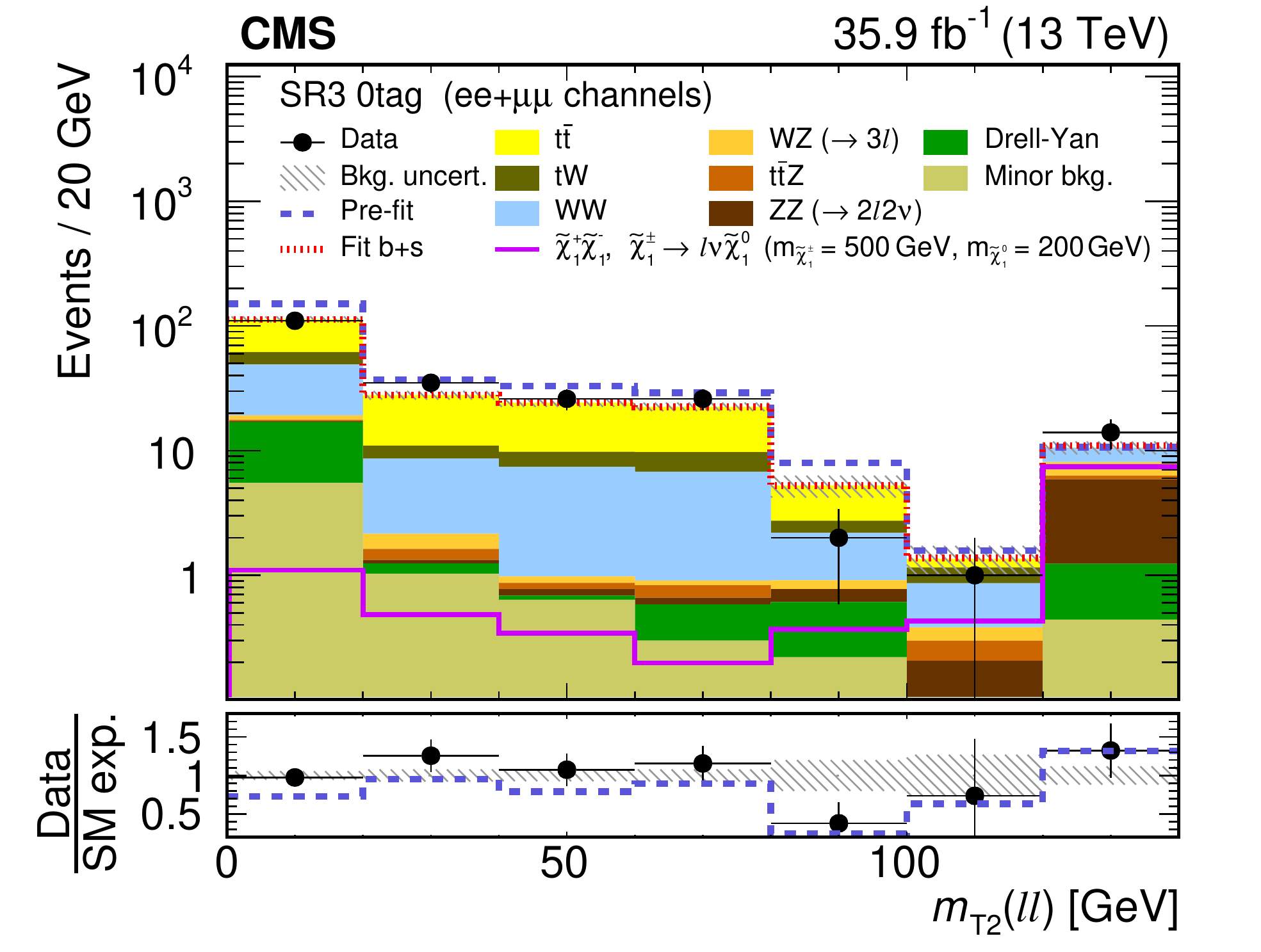}
\caption{The same distributions of \mtll as Fig.~\ref{Fig:MT2ll_TChi_em}, but for SF events.}
\label{Fig:MT2ll_TChi_sf}
\end{figure}

\begin{figure}
\centering
\includegraphics[width=0.48\textwidth]{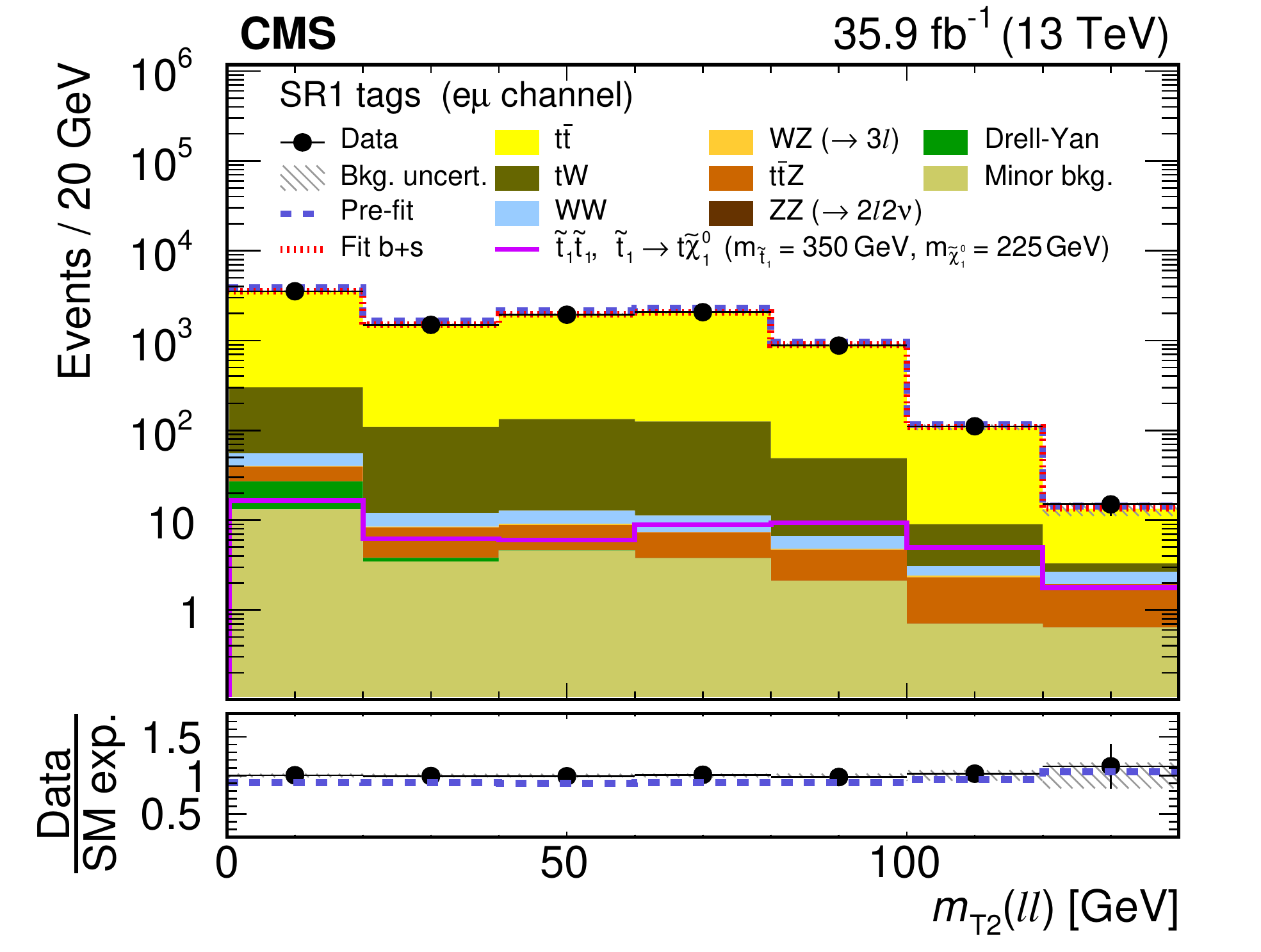}
\includegraphics[width=0.48\textwidth]{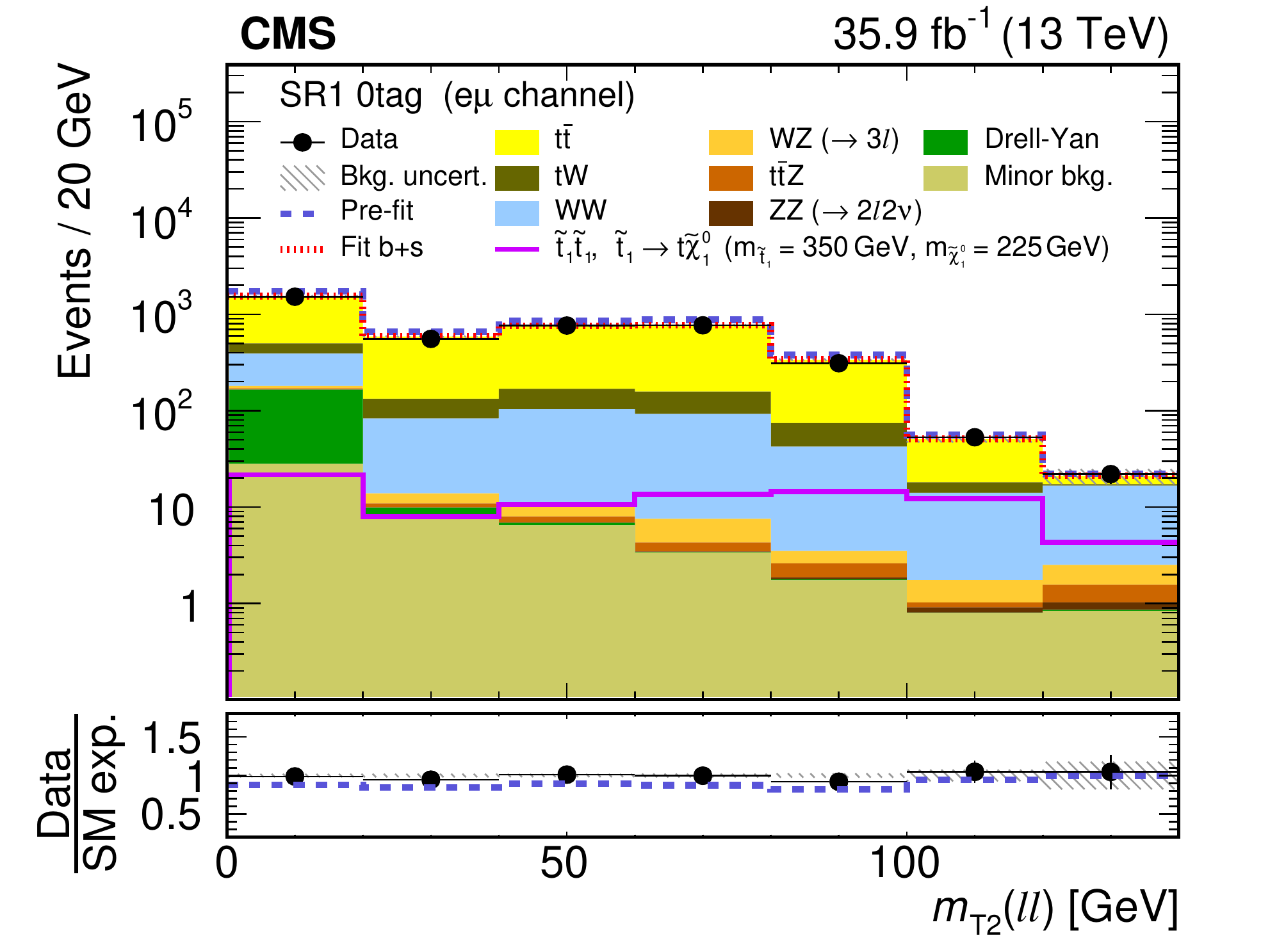}
\includegraphics[width=0.48\textwidth]{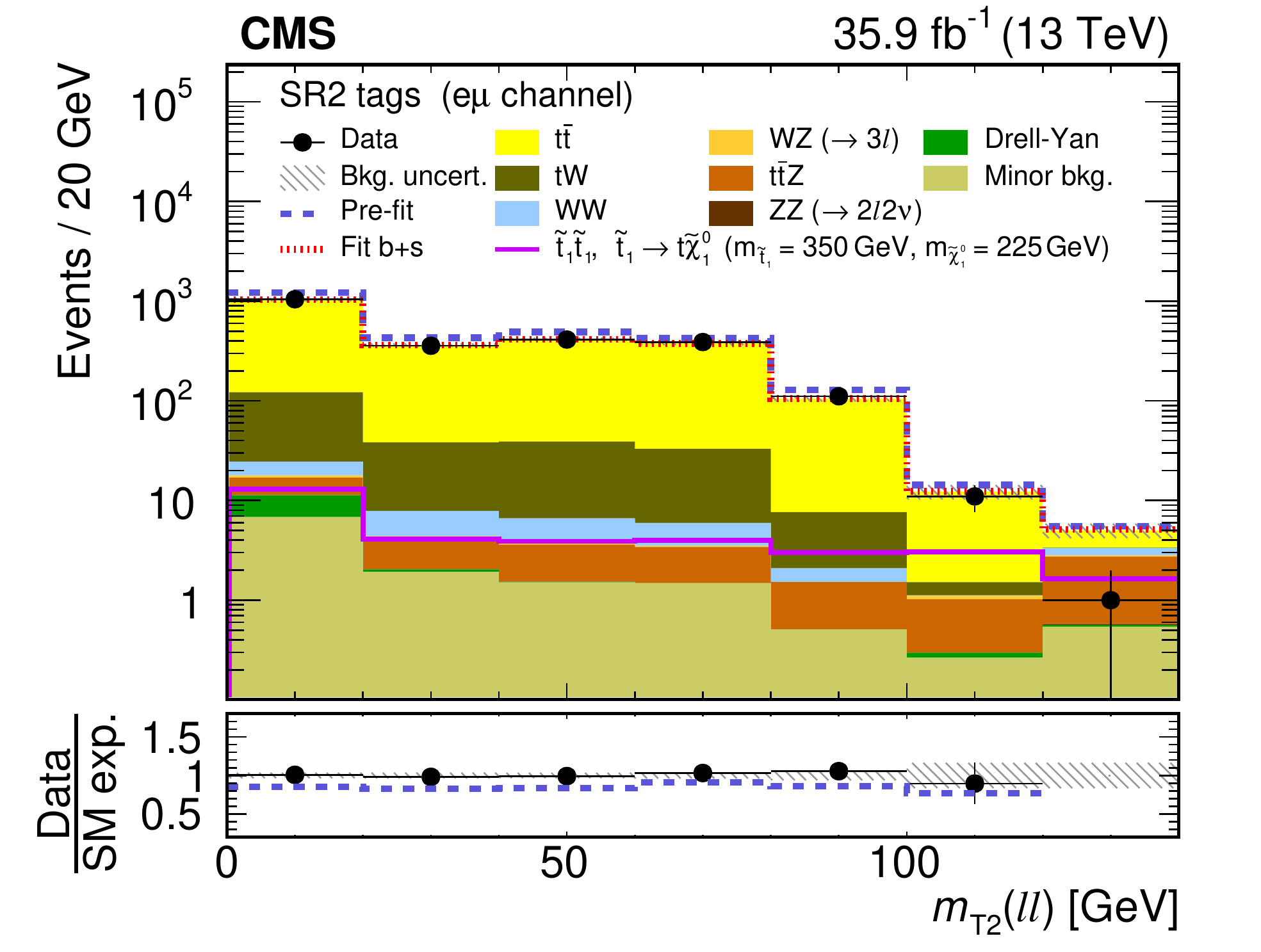}
\includegraphics[width=0.48\textwidth]{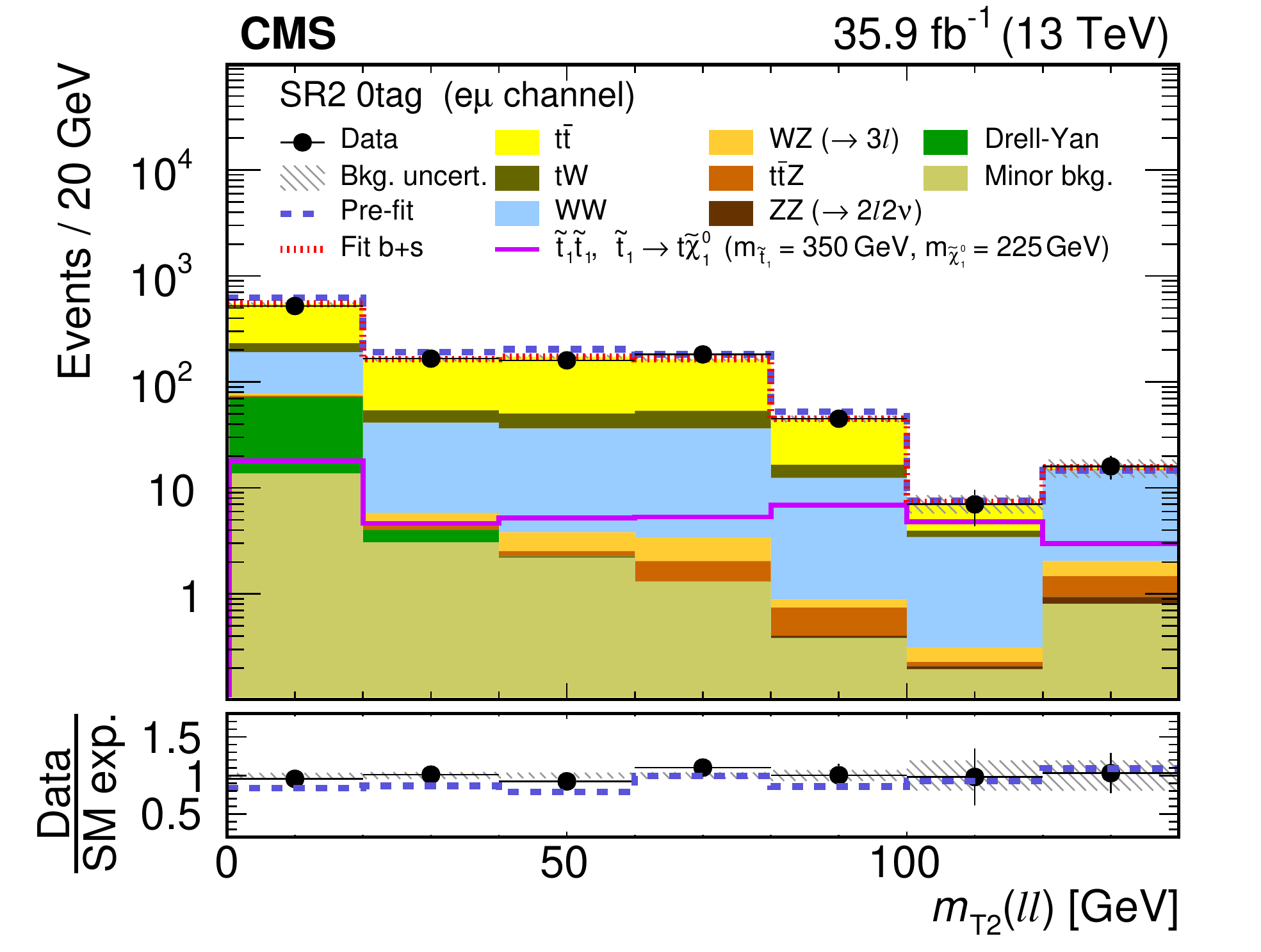}
\includegraphics[width=0.48\textwidth]{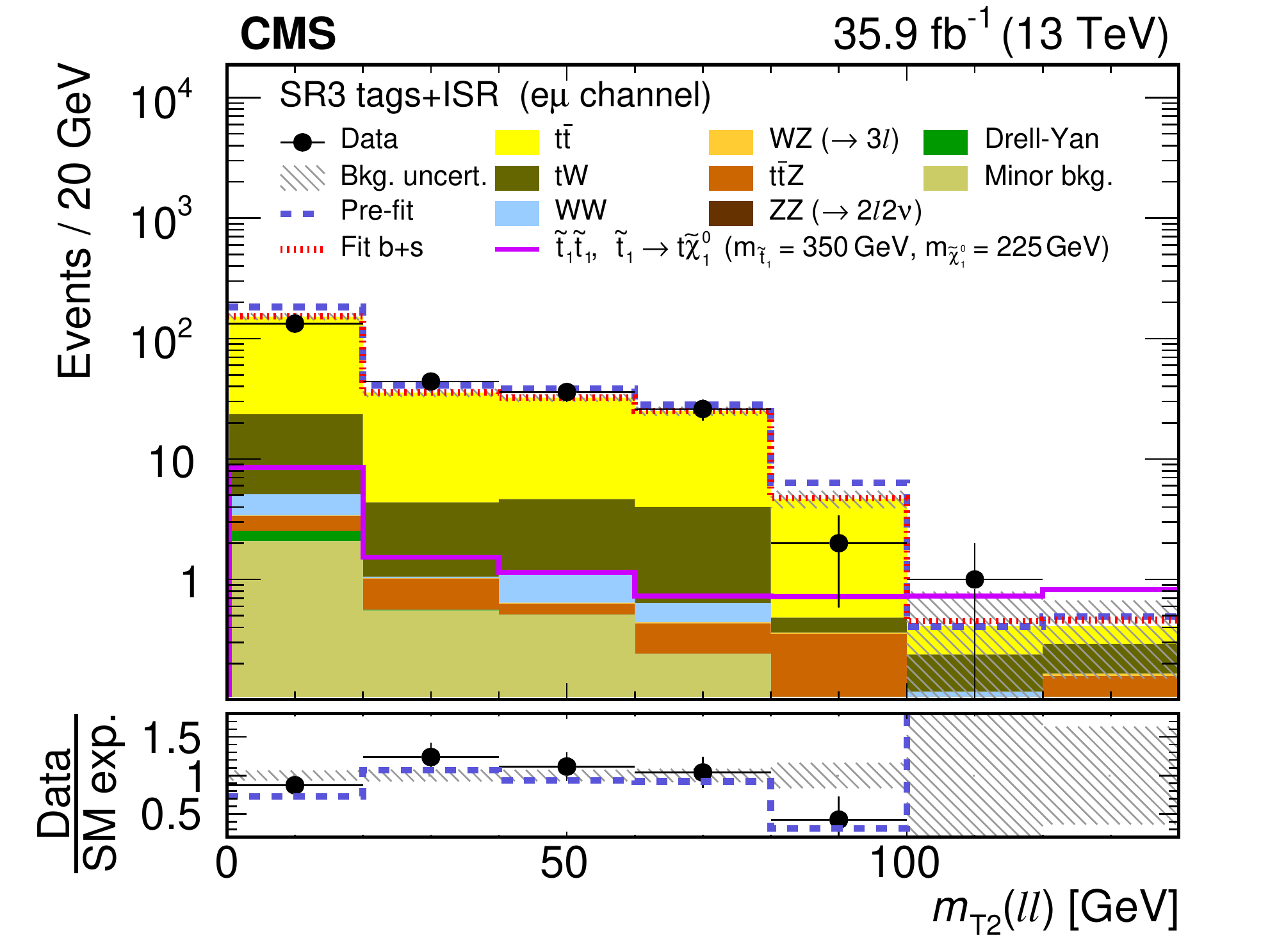}
\includegraphics[width=0.48\textwidth]{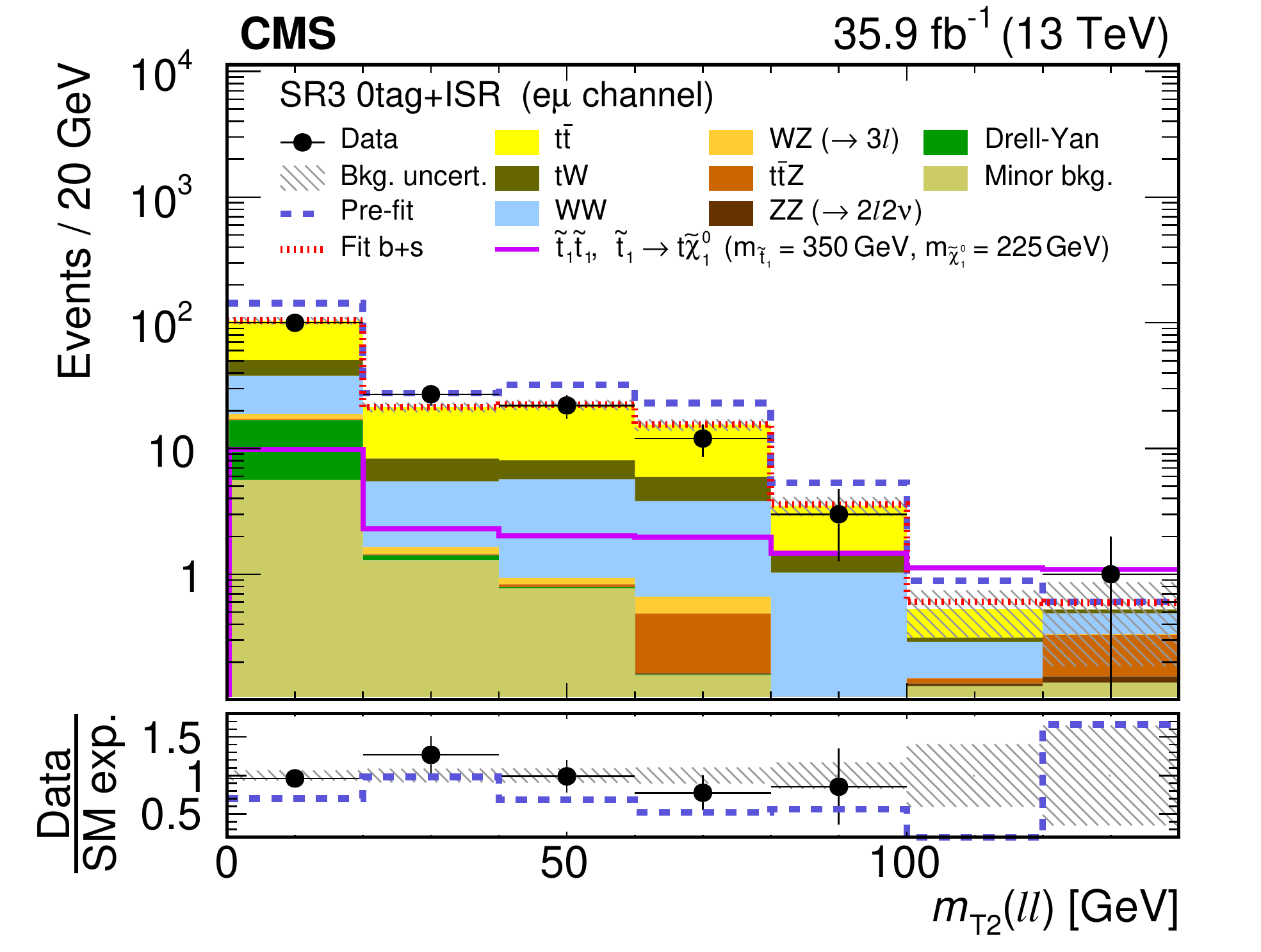}
\caption{Distributions of \mtll after the fit to data in the top squark SRs with $140<\ptmiss<200\GeV$ (upper plots),  $200<\ptmiss<300\GeV$ (middle), or $\ptmiss>300\GeV$ (lower), for DF events  with \cPqb-tagged jets (left plots) and without \cPqb-tagged jets (right plots). The solid magenta histogram shows the expected \mtll distribution for top squark pair production with $\invM{\stone}=350\GeV$ and $\invM{\PSGczDo}=225\GeV$. Expected total SM contributions before the fit (dark blue dashed line) and after a background+signal fit (dark red dotted line) are also shown. The last bin includes the overflow entries. In the bottom panel, the ratio of data and SM expectations is shown for the expected total SM contribution after the fit using the background-only hypothesis (black dots) and before any fit (dark blue dashed line). The hatched band represents the total uncertainty after the fit.}
\label{Fig:MT2ll_T2_em}
\end{figure}

\begin{figure}
  \centering
  \includegraphics[width=0.48\textwidth]{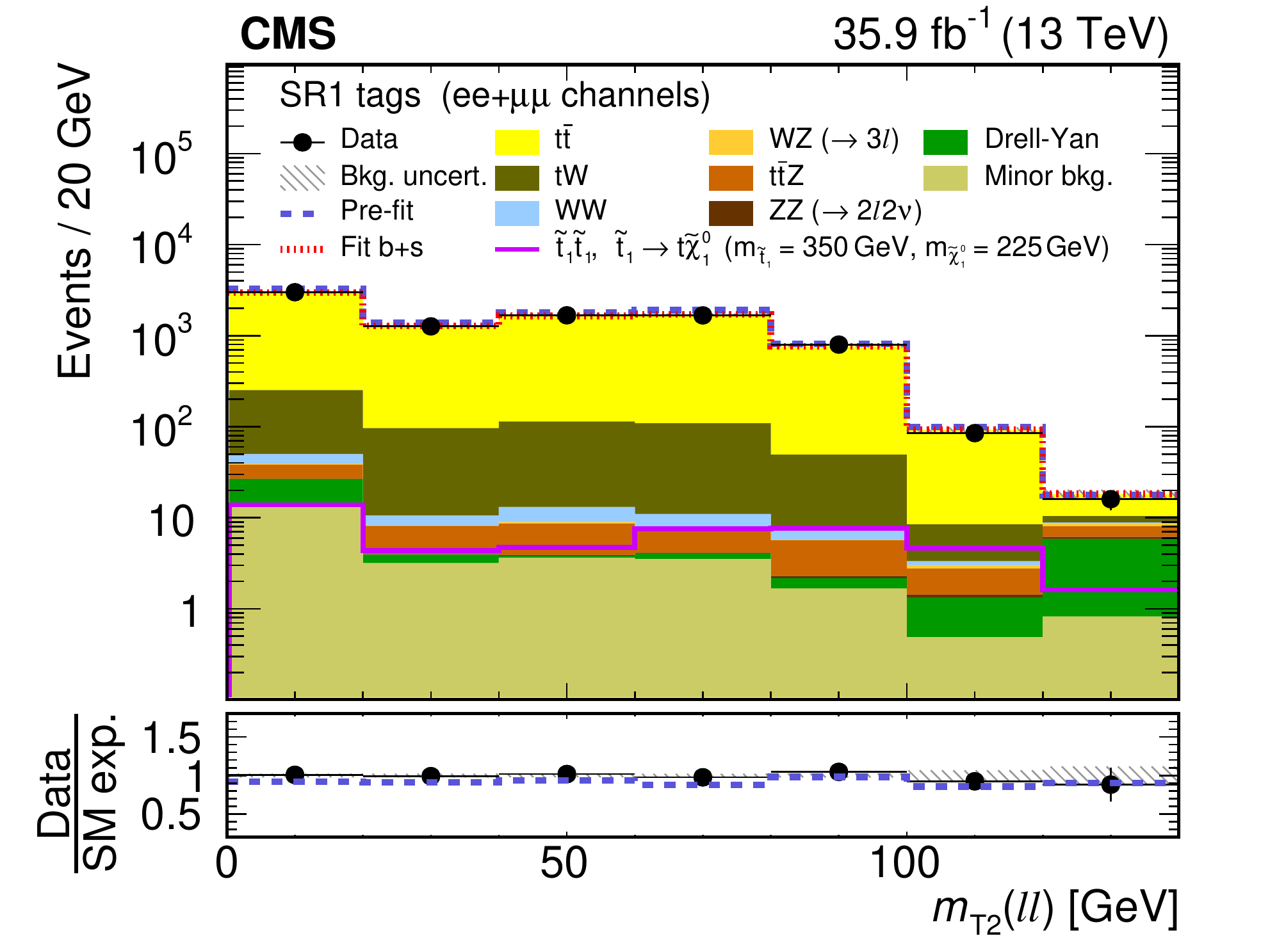}
  \includegraphics[width=0.48\textwidth]{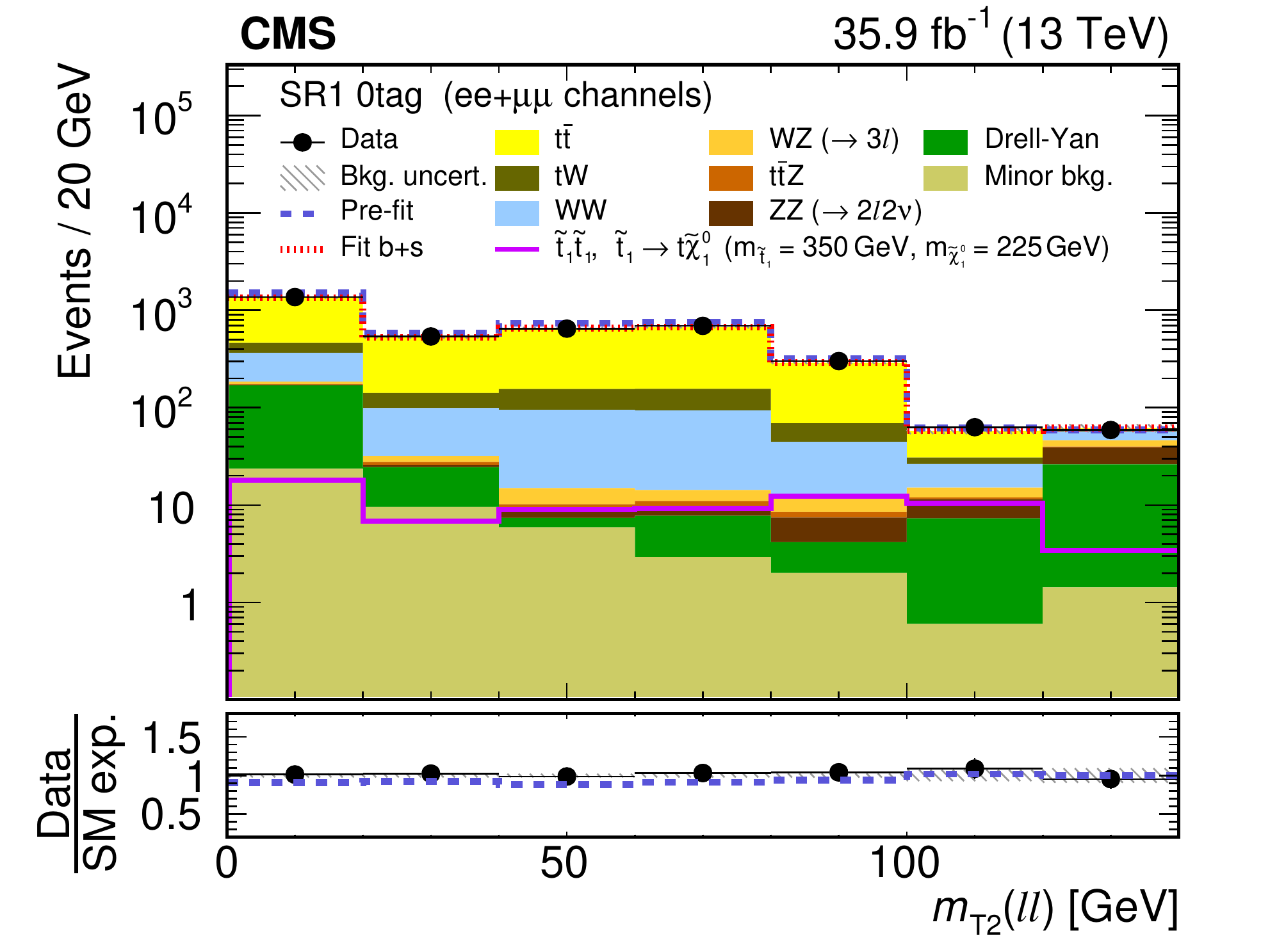}
  \includegraphics[width=0.48\textwidth]{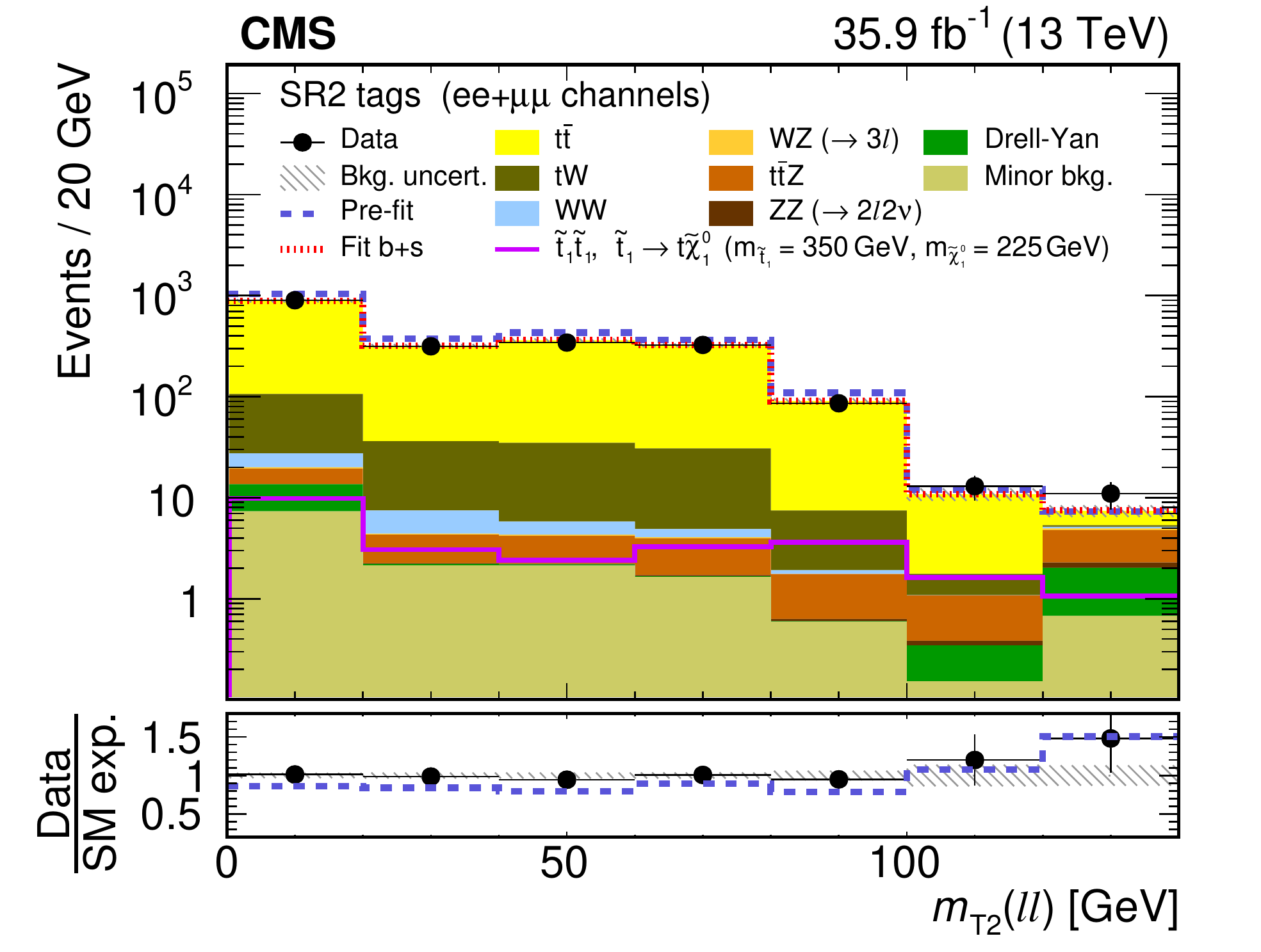}
  \includegraphics[width=0.48\textwidth]{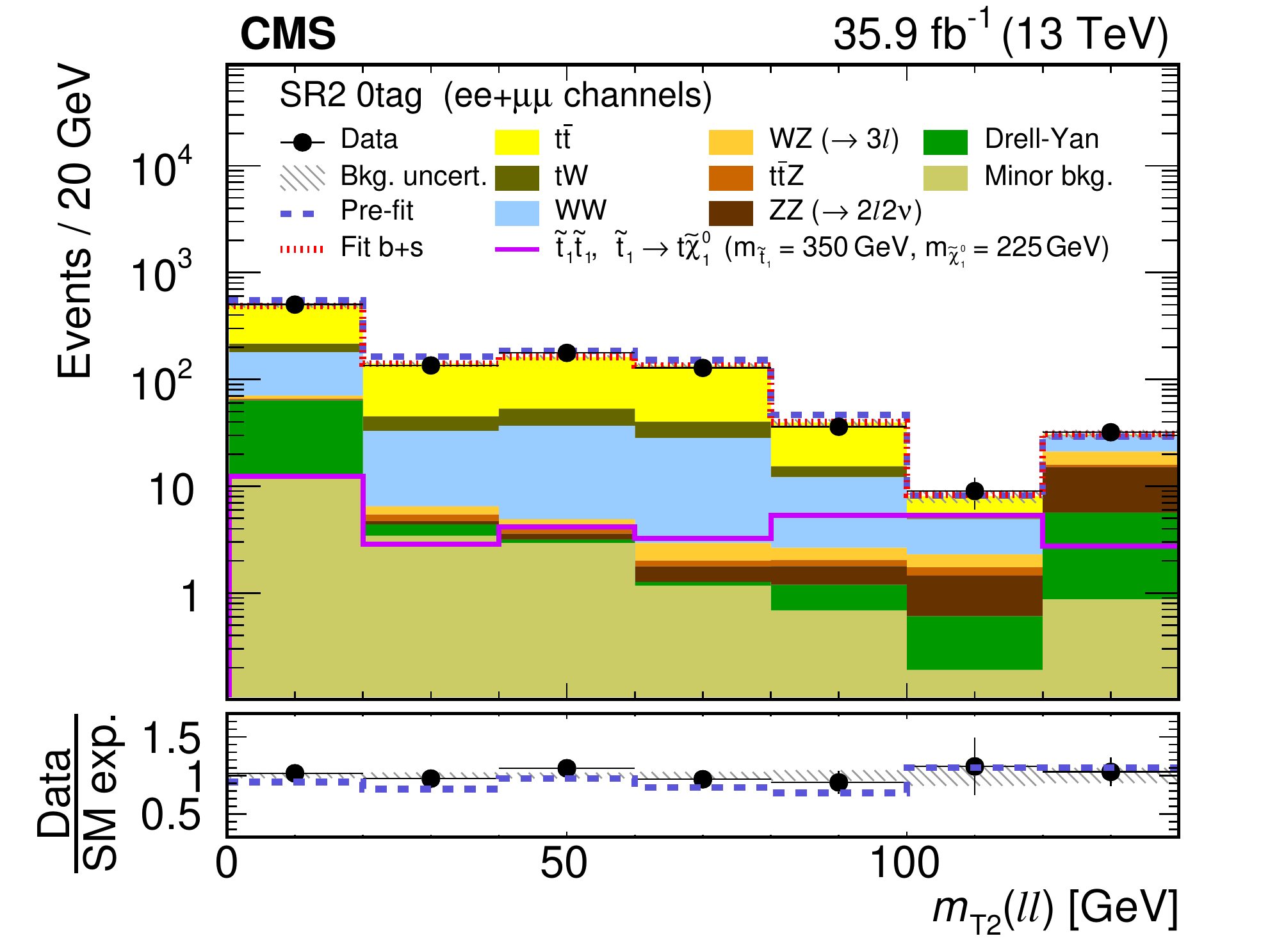}
  \includegraphics[width=0.48\textwidth]{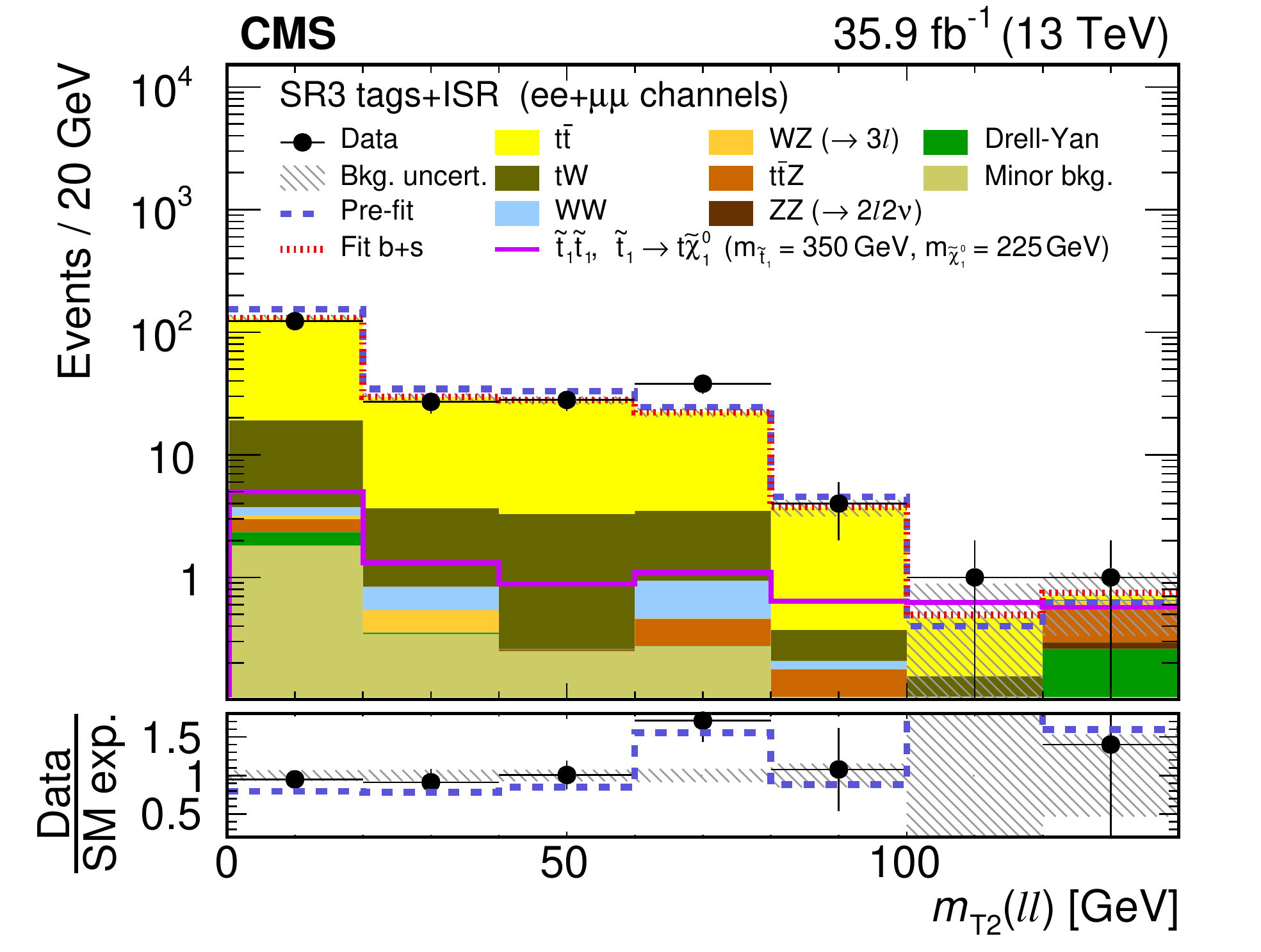}
  \includegraphics[width=0.48\textwidth]{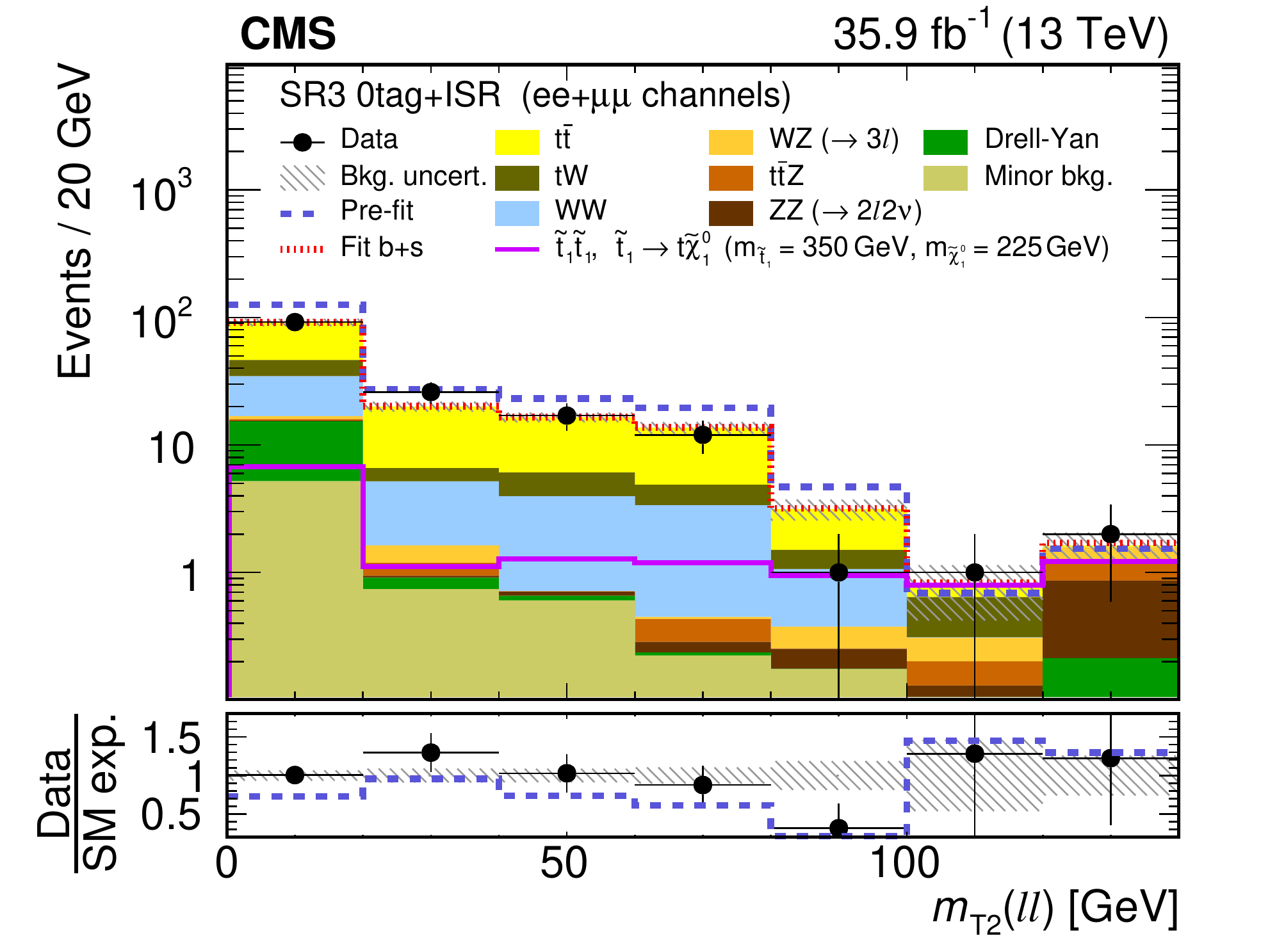}
    \caption{The same distributions of \mtll as Fig.~\ref{Fig:MT2ll_T2_em}, but for SF events.}
   \label{Fig:MT2ll_T2_sf}
\end{figure}

A simultaneous binned ML fit to the \mtll distribution in all the SRs is performed.
Uncertainties due to signal and background normalizations are included through nuisance parameters with log-normal prior distributions, while uncertainties in the shape of the \mtll distributions are included with Gaussian prior distributions. As explained in Section~\ref{sec:bkgestimation}, the normalizations of the main backgrounds from top quark and \WW production are left to be determined in the fit via the constraint provided by the low \mtll region with and without \cPqb-tagged jets.
The results of the fit in the SRs for the chargino search are shown in Figs.~\ref{Fig:MT2ll_TChi_em} and~\ref{Fig:MT2ll_TChi_sf} for DF and SF events, respectively. The results for the top squark search are shown in Figs.~\ref{Fig:MT2ll_T2_em} and~\ref{Fig:MT2ll_T2_sf}.
Each figure compares the number of observed events in the SRs with the expected yields from SM processes after a background-only fit. As a comparison, the expected yields for a representative signal point are given. The total expected SM contributions before the fit and after a background+signal fit are also shown.
Detailed information on the observed and expected yields after the background-only fit are given in Tables~\ref{tab:PostfitTChill_Short}--\ref{tab:PostfitT2ttll_Short} for all dilepton final states and all SRs.
No excess over SM prediction is observed in data.
The asymptotic approximation of the \CLs criterion~\cite{bib:CL2,bib:CLs,bib:AsymptCL} is used to set upper limits at 95\% confidence level (\CL) on the production cross sections for the different signal models considered.

\begin{table}[htb]
\centering
\topcaption{Observed and expected yields of DF (the upper half of Table) and SF (the lower half) events in the SRs for the chargino search. The quoted uncertainties in the background predictions include statistical and systematic contributions.}
\cmsTable{
\begin{tabular}{llccccccc}
\hline
\multicolumn{2}{c}{\mtll [{\GeVns}]} &0--20 & 20--40 & 40--60 & 60--80 & 80--100 & 100--120 & ${\ge}120$ \\
\hline
DF events  & &  &  & & & &  \\[\cmsTabSkip]
\multirow{2}{*}{\Reg{SR1}{0tag}{jets}} &  Predicted  &  $1493 \pm 32$ &  $558 \pm 12$ &  $719 \pm 16$ &  $730 \pm 16$ &  $316 \pm 10$ &  $45.1 \pm 3.1$ &  $13.7 \pm 2.8$ \\
                             &  Observed  & 1484 & 532 & 732 & 725 & 298 & 47 & 13 \\[\cmsTabSkip]
\multirow{2}{*}{\Reg{SR1}{0tag}{0jet}} & Predicted  &  $41.9 \pm 5$ &  $27.4 \pm 3.8$ &  $34.1 \pm 4.8$ &  $42 \pm 5.5$ &  $21.1 \pm 3.4$ &  $6 \pm 1.3$ &  $7.9 \pm 2.1$ \\
                             & Observed  & 39 & 24 & 33 & 44 & 13 & 6 & 9 \\[\cmsTabSkip]
\multirow{2}{*}{\Reg{SR2}{0tag}{jets}} & Predicted  &  $534 \pm 15$ &  $158.6 \pm 5.9$ &  $167.9 \pm 6.1$ &  $157.9 \pm 6.5$ &  $42.4 \pm 2.9$ &  $5.9 \pm 1$ &  $9 \pm 1.7$ \\
                             & Observed  & 511 & 162 & 156 & 176 & 43 & 5 & 9 \\[\cmsTabSkip]
\multirow{2}{*}{\Reg{SR2}{0tag}{0jet}} & Predicted  &  $10.3 \pm 1.7$ &  $7 \pm 1.5$ &  $6.5 \pm 1.3$ &  $6.9 \pm 1.3$ &  $2.19 \pm 0.69$ &  $1.59 \pm 0.7$ &  $7.8 \pm 1.8$ \\
                             & Observed  & 10 & 4 & 4 & 6 & 2 & 2 & 7 \\[\cmsTabSkip]
\multirow{2}{*}{\Reg{SR3}{0tag}{}} & Predicted  &  $127.9 \pm 7.2$ &  $28.3 \pm 2$ &  $30.2 \pm 2.4$ &  $23.1 \pm 2$ &  $4.96 \pm 0.73$ &  $1.12 \pm 0.38$ &  $4.5 \pm 1.2$ \\
                  & Observed  & 116 &        35            &  29             & 21           & 3             & 1               & 5 \\[\cmsTabSkip]
SF events  & &  &  & & & &  \\[\cmsTabSkip]
\multirow{2}{*}{\Reg{SR1}{0tag}{jets}} & Predicted  &  $1310 \pm 29$ &  $499 \pm 12$ &  $623 \pm 14$ &  $634 \pm 15$ &  $271.7 \pm 8.9$ &  $51.6 \pm 3.5$ &  $48.6 \pm 5.5$ \\
                              & Observed  & 1324 & 499 & 609 & 659 & 284 & 57 & 47 \\[\cmsTabSkip]
\multirow{2}{*}{\Reg{SR1}{0tag}{0jet}} & Predicted  &  $44.1 \pm 7.5$ &  $28.5 \pm 4.1$ &  $33.5 \pm 4.4$ &  $33.5 \pm 4.5$ &  $18.6 \pm 2.6$ &  $7.7 \pm 1.6$ &  $12.5 \pm 2.5$ \\
                              & Observed  & 43 & 40 & 39 & 33 & 17 & 6 & 12 \\[\cmsTabSkip]
\multirow{2}{*}{\Reg{SR2}{0tag}{jets}} & Predicted  &  $474 \pm 14$ &  $134.8 \pm 5.1$ &  $155.1 \pm 5.5$ &  $128.5 \pm 5.5$ &  $37.1 \pm 2.5$ &  $7.29 \pm 0.91$ &  $23.9 \pm 2.4$ \\
                              & Observed  & 493 & 123 & 166 & 118 & 33 & 7 & 25 \\[\cmsTabSkip]
\multirow{2}{*}{\Reg{SR2}{0tag}{0jet}} & Predicted  &  $10.9 \pm 1.9$ &  $7.8 \pm 1.8$ &  $7.3 \pm 1.4$ &  $7.9 \pm 1.3$ &  $1.9 \pm 0.52$ &  $1.28 \pm 0.58$ &  $7.1 \pm 1.4$ \\
                              & Observed  & 8 & 12 & 11 & 10 & 3 & 2 & 7 \\[\cmsTabSkip]
\multirow{2}{*}{\Reg{SR3}{0tag}{}} & Predicted  &  $112.8 \pm 6.3$ &  $27.9 \pm 2.2$ &  $24.2 \pm 1.8$ &  $22.5 \pm 1.8$ &  $5.2 \pm 1$ &  $1.36 \pm 0.36$ &  $10.6 \pm 1.2$ \\
                  & Observed  & 110 & 35 & 26 & 26 & 2 & 1 & 14 \\
\hline
\end{tabular}
}
\label{tab:PostfitTChill_Short}
\end{table}

\begin{table}[htb]
\centering
\topcaption{Observed and expected yields of DF (the upper half of Table) and SF (the lower half) events in the SRs for the top squark search. The quoted uncertainties in the background predictions include statistical and systematic contributions.}
\cmsTable{
\begin{tabular}{llccccccc}
\hline
 \multicolumn{2}{c}{\mtll [{\GeVns}]} &0--20 & 20--40 & 40--60 & 60--80 & 80--100 & 100--120 & ${\ge}120$ \\
\hline
DF events & &  &  & & & &  \\[\cmsTabSkip]
\multirow{2}{*}{\Reg{SR1}{tags}{}} &  Predicted  &  $3525 \pm 80$ &  $1505 \pm 31$ &  $1958 \pm 42$ &  $2049 \pm 46$ &  $897 \pm 22$ &  $108.4 \pm 7.3$ &  $13.4 \pm 2.2$ \\
                  &  Observed  & 3534 & 1494 & 1938 & 2068 & 879 & 111 & 15 \\[\cmsTabSkip]
\multirow{2}{*}{\Reg{SR1}{0tag}{}} & Predicted  &  $1542 \pm 33$ &  $588 \pm 13$ &  $756 \pm 15$ &  $771 \pm 19$ &  $338.3 \pm 9.3$ &  $50.6 \pm 3.8$ &  $21 \pm 3.8$ \\
                  & Observed  & 1523 & 556 & 765 & 769 & 311 & 53 & 22 \\[\cmsTabSkip]
\multirow{2}{*}{\Reg{SR2}{tags}{}} & Predicted  &  $1036 \pm 37$ &  $363 \pm 13$ &  $415 \pm 14$ &  $377 \pm 14$ &  $105.1 \pm 6.5$ &  $12.3 \pm 2$ &  $5.02 \pm 0.82$ \\
                  & Observed  & 1045 & 357 & 412 & 389 & 111 & 11 & 1 \\[\cmsTabSkip]
\multirow{2}{*}{\Reg{SR2}{0tag}{}} & Predicted  &  $545 \pm 18$ &  $164.3 \pm 7.3$ &  $173.2 \pm 6.2$ &  $165.1 \pm 6.8$ &  $44.8 \pm 3.1$ &  $7.1 \pm 1.4$ &  $15.5 \pm 3$ \\
                  & Observed  & 521 & 166 & 160 & 182 & 45 & 7 & 16 \\[\cmsTabSkip]
\multirow{2}{*}{\Reg{SR3}{tags}{ISR}} & Predicted  &  $152.1 \pm 9.9$ &  $35.5 \pm 2.7$ &  $32.3 \pm 2.3$ &  $25 \pm 2.2$ &  $4.67 \pm 0.77$ &  $0.41 \pm 0.38$ &  $0.41 \pm 0.26$ \\
                             & Observed  & 133 & 44 & 36 & 26 & 2 & 1 & 0 \\[\cmsTabSkip]
\multirow{2}{*}{\Reg{SR3}{0tag}{ISR}} & Predicted  &  $103.9 \pm 6.8$ &  $21.3 \pm 1.9$ &  $22.2 \pm 2.1$ &  $15.4 \pm 1.6$ &  $3.51 \pm 0.6$ &  $0.53 \pm 0.21$ &  $0.53 \pm 0.34$ \\
                             & Observed  & 100 & 27 & 22 & 12 & 3 & 0 & 1 \\[\cmsTabSkip]
SF events  & &  &  & & & &  \\[\cmsTabSkip]
\multirow{2}{*}{\Reg{SR1}{tags}{}} &  Predicted  &  $2979 \pm 68$ &  $1277 \pm 30$ &  $1644 \pm 35$ &  $1712 \pm 37$ &  $762 \pm 19$ &  $91.9 \pm 6.1$ &  $18.1 \pm 2.1$ \\
                  &  Observed  & 3003 & 1266 & 1674 & 1671 & 798 & 85 & 16 \\[\cmsTabSkip]
\multirow{2}{*}{\Reg{SR1}{0tag}{}} & Predicted  &  $1350 \pm 33$ &  $526 \pm 13$ &  $656 \pm 15$ &  $670 \pm 17$ &  $289.2 \pm 7.6$ &  $57.9 \pm 4.2$ &  $61.8 \pm 5.8$ \\
                  & Observed  & 1367 & 539 & 648 & 692 & 301 & 63 & 59 \\[\cmsTabSkip]
\multirow{2}{*}{\Reg{SR2}{tags}{}} & Predicted  &  $888 \pm 30$ &  $319 \pm 12$ &  $363 \pm 14$ &  $323 \pm 13$ &  $90.5 \pm 5.5$ &  $10.8 \pm 1.5$ &  $7.43 \pm 0.98$ \\
                  & Observed  & 900 & 315 & 343 & 325 & 86 & 13 & 11 \\[\cmsTabSkip]
\multirow{2}{*}{\Reg{SR2}{0tag}{}} & Predicted  &  $487 \pm 16$ &  $140.7 \pm 5.5$ &  $161.9 \pm 5.9$ &  $134.5 \pm 6.2$ &  $39.6 \pm 2.7$ &  $8.1 \pm 1.1$ &  $30.6 \pm 3$ \\
                  & Observed  & 501 & 135 & 177 & 128 & 36 & 9 & 32 \\[\cmsTabSkip]
\multirow{2}{*}{\Reg{SR3}{tags}{ISR}} & Predicted  &  $129.6 \pm 8.9$ &  $29.6 \pm 2.1$ &  $27.8 \pm 2.1$ &  $22.2 \pm 1.9$ &  $3.71 \pm 0.57$ &  $0.47 \pm 0.42$ &  $0.71 \pm 0.38$ \\
                             & Observed  & 123 & 27 & 28 & 38 & 4 & 1 & 1 \\[\cmsTabSkip]
 \multirow{2}{*}{\Reg{SR3}{0tag}{ISR}} & Predicted  &  $91.5 \pm 6.1$ &  $20.1 \pm 1.8$ &  $16.5 \pm 1.4$ &  $13.7 \pm 1.4$ &  $3.14 \pm 0.58$ &  $0.78 \pm 0.36$ &  $1.63 \pm 0.42$ \\
                              & Observed  & 92 & 26 & 17 & 12 & 1 & 1 & 2 \\
\hline
\end{tabular}
}
\label{tab:PostfitT2ttll_Short}
\end{table}

\begin{figure}[hbtp]
  \centering
  \includegraphics[width=0.48\textwidth]{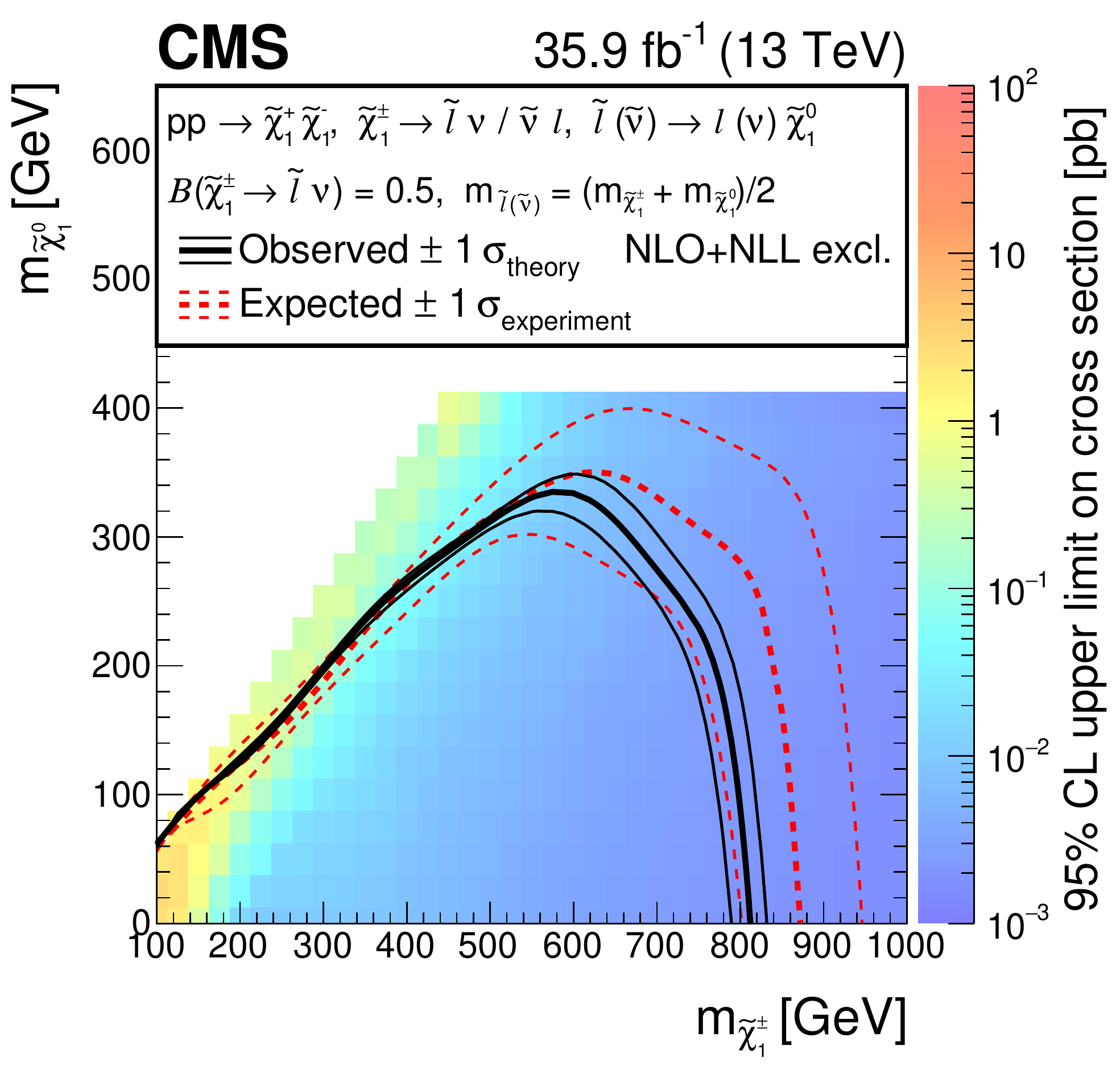}
  \includegraphics[width=0.48\textwidth]{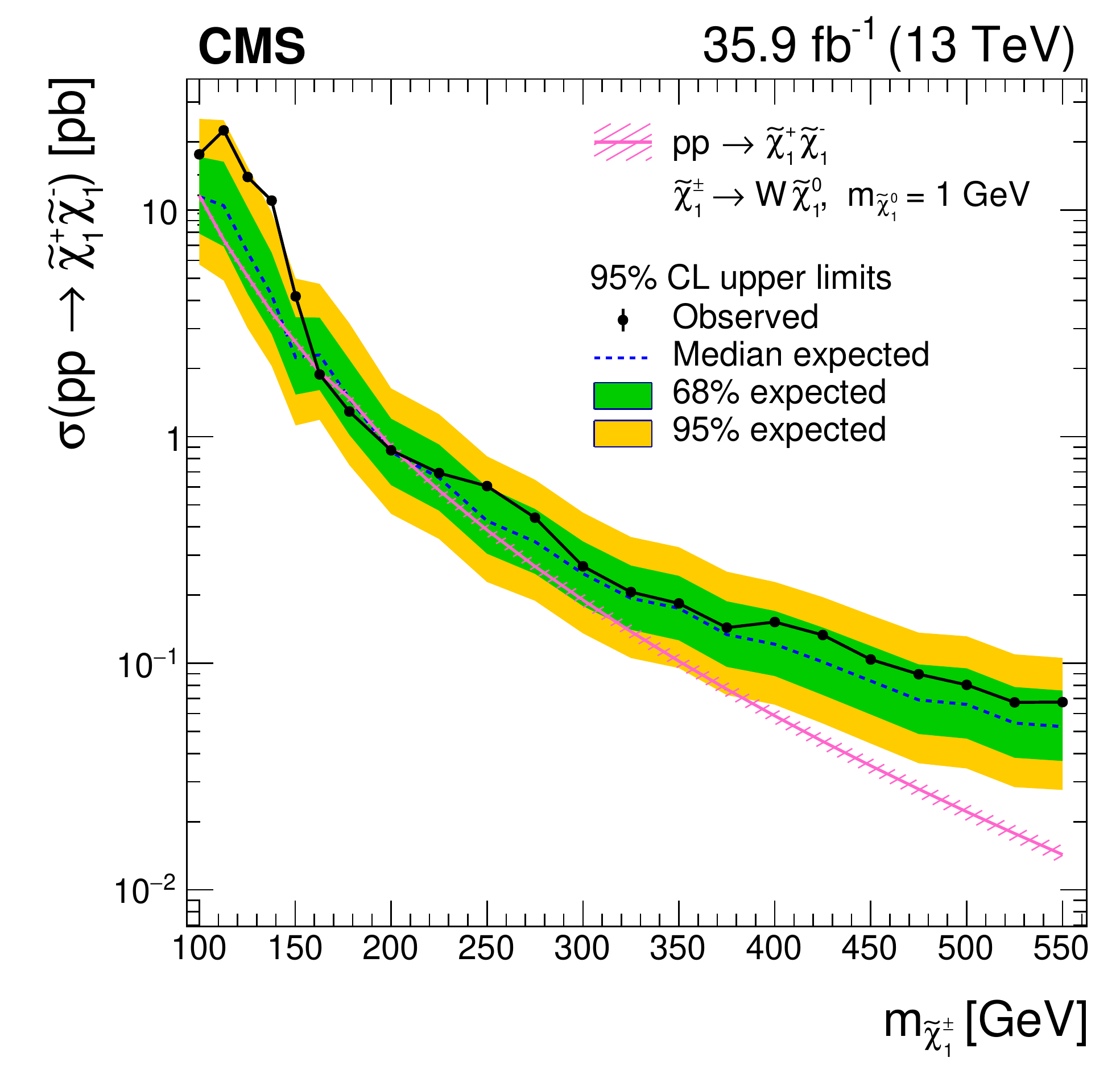}
    \caption{Left: upper limits at 95\% \CL on chargino pair production cross section as a function of the chargino and neutralino masses, when the chargino undergoes a cascade decay \TChipmDecay. Exclusion regions in the plane (\invM{\PSGcpmDo}, \invM{\PSGczDo}) are determined by comparing the upper limits with the NLO+NLL production cross sections. The thick dashed red line shows the expected exclusion region. The thin dashed red lines show the variation of the exclusion regions due to the experimental uncertainties. The thick black line shows the observed exclusion region, while the thin black lines show the variation of the exclusion regions due to the theoretical uncertainties in the production cross section. Right: observed and expected upper limits at 95\% \CL as a function of the chargino mass for a neutralino mass of 1\GeV, assuming chargino decays into a neutralino and a {\PW} boson ($\PSGcpmDo\to\PW\PSGczDo$).}
    \label{Fig:TChipmLimit}
\end{figure}

The 95\% \CL upper limits on  chargino pair production cross sections with the chargino decaying into sleptons are shown in Fig.~\ref{Fig:TChipmLimit} (left). The \TChipmSlep and \TChipmSneu decay chains are given a \BF of 50\% each, and the sleptons are assumed to be degenerate, with a mass equal to the average of the chargino and neutralino masses. By comparing the upper limits with $\pp\to\PSGcpDo\PSGcmDo$ production cross sections, observed and expected exclusion regions in the (\invM{\PSGcpmDo}, \invM{\PSGczDo}) plane are also determined.
Masses are excluded up to values of about 800 and 320\GeV for the chargino and the neutralino, respectively.
Limited sensitivity is found when the chargino is assumed to decay into a {\PW} boson and the lightest neutralino, due to the relatively small \BF for the leptonic decay of the {\PW} boson. For this scenario, we derive upper limits on chargino pair production cross section assuming a lightest neutralino mass of 1\GeV. Observed and expected upper limits as a function of the chargino mass are compared to theoretical cross sections in Fig.~\ref{Fig:TChipmLimit} (right).

\begin{figure}[hbtp]
  \centering
  \includegraphics[width=0.48\textwidth]{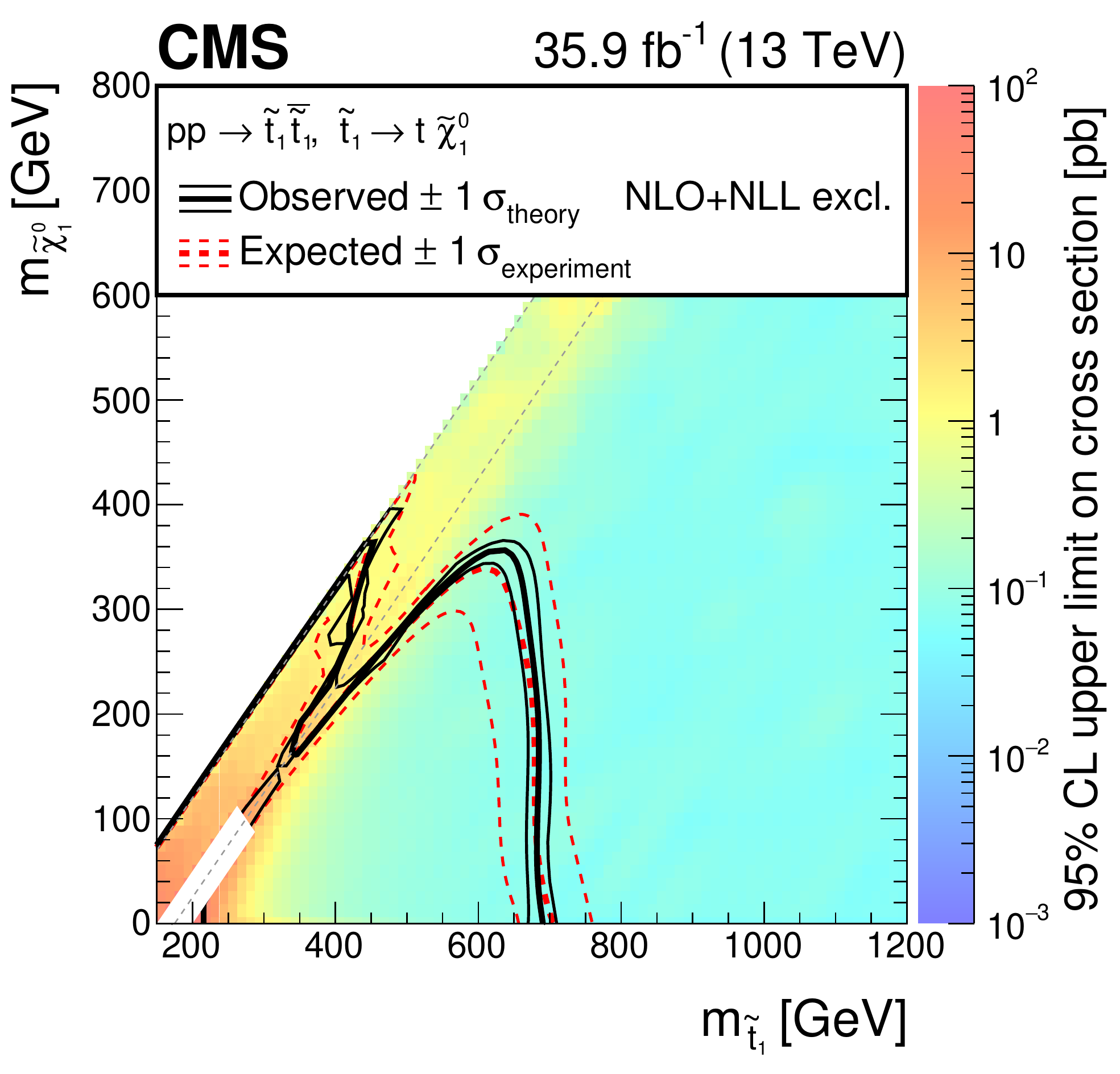}
  \includegraphics[width=0.48\textwidth]{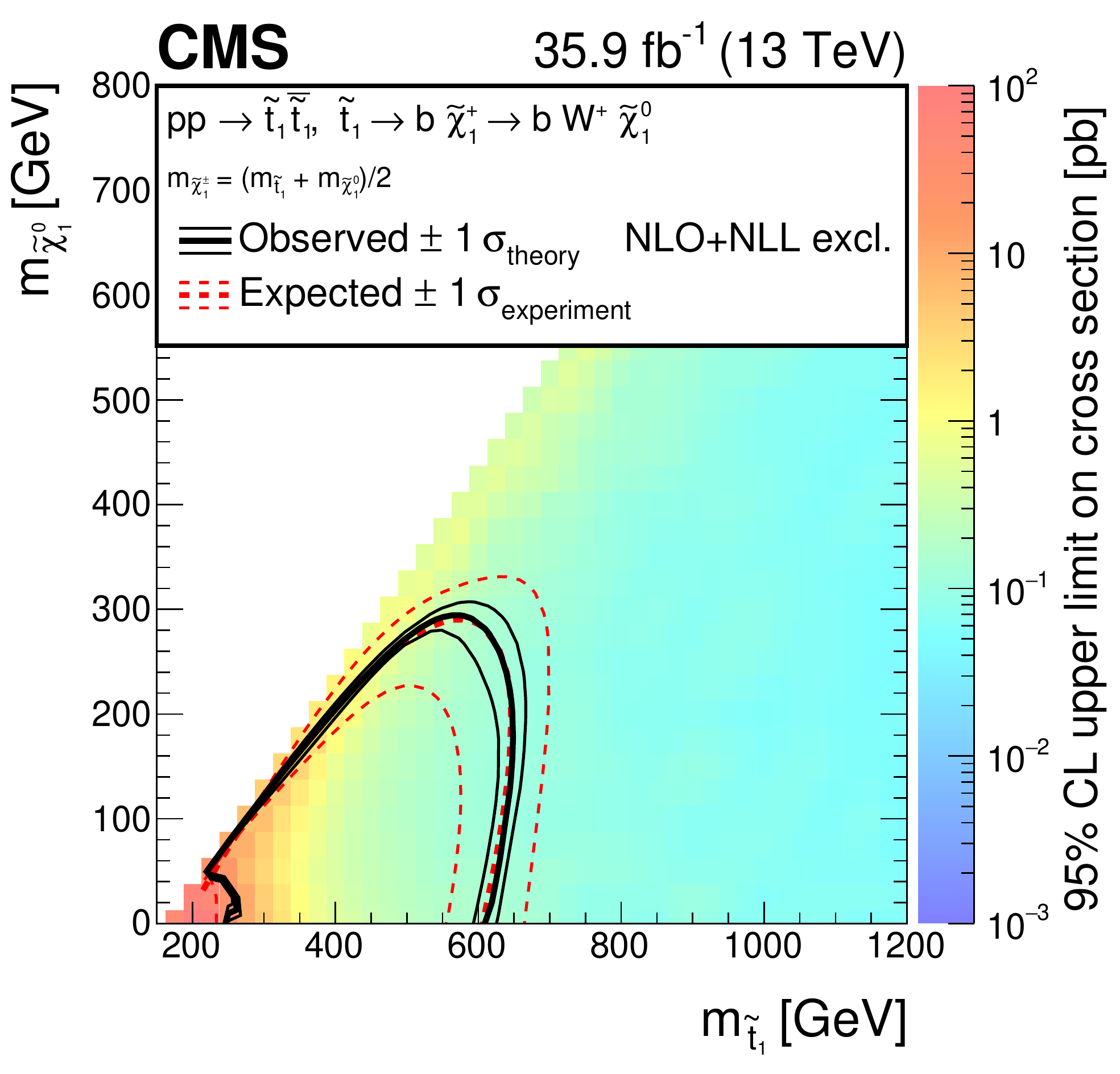}
    \caption{Upper limits at 95\% \CL on top squark production cross section as a function of the top squark and neutralino masses. The plot on the left shows the results when top squark decays into a top quark and a neutralino are assumed. The two diagonal gray dashed lines enclose the compressed region where $\invM{\PW}<\invM{\stone}-\invM{\PSGczDo}\lesssim \invM{\PQt}$. The plot on the right gives the limits for top squarks decaying into a bottom quark and a chargino, with the latter successively decaying into a {\PW} boson and a neutralino. The mass of the chargino is assumed to be equal to the average of the top squark and neutralino masses. Exclusion regions in the plane (\invM{\stone}, \invM{\PSGczDo}) are determined by comparing the upper limits with the NLO+NLL production cross sections. The thick dashed red line shows the expected exclusion region. The thin dashed red lines show the variation of the exclusion regions due to the experimental uncertainties. The thick black line shows the observed exclusion region, while the thin black lines show the variation of the exclusion regions due to the theoretical uncertainties in the production cross section.}
    \label{Fig:T2Limit}
\end{figure}

Figure~\ref{Fig:T2Limit} shows the observed and expected 95\% \CL upper limits on top squark production cross section for the two SMS considered. While the search strategy has been optimized for a compressed scenario, the results are presented on the whole (\invM{\stone}, \invM{\PSGczDo}) plane for completeness.
Also shown are the expected and observed exclusion regions when assuming NLO+NLL top squark pair production cross sections.
When assuming the top squark to decay into a top quark and a neutralino, top squark (neutralino) masses are excluded up to about 420 (360)\GeV in the compressed mass region where \Dm lies between the top quark and {\PW} boson masses.
For the $\,\stone\to \cPqb\PSGcpmDo\to \cPqb\PW\PSGczDo$ decay mode, a lower bound $\Dm\approx 2\,\invM{\PW}$ is set by the assumption that $\invM{\PSGcpmDo}=(\invM{\stone}+\invM{\PSGczDo})/2$. For $\Dm\approx 2\,\invM{\PW}$, top squark masses are excluded in the range 225--325\GeV.
The uncovered region around a top squark mass of 200\GeV in Fig.~\ref{Fig:T2Limit} (right) corresponds to a signal phase space similar to that of \ttbar events, with little contribution from the neutralinos to \ptvecmiss. In this situation, the uncertainty in the modeling of \ptvecmiss in \FastSim events becomes too large to provide any signal sensitivity.

\section{Summary}\label{sec:summary}

A search has been presented for pair production of supersymmetric particles in events with two oppositely charged isolated leptons and missing transverse momentum.
The data used consist of a sample of proton-proton collisions collected with the CMS detector during the 2016 LHC run at a center-of-mass energy of 13\TeV, corresponding to an integrated luminosity of \Lum.
No evidence for a deviation with respect to standard model predictions was observed in data. The results have been interpreted as upper limits on the cross sections of supersymmetric particle production for several simplified model spectra.

Chargino pair production has been investigated in two possible decay modes.
If the chargino is assumed to undergo a cascade decay through sleptons, an exclusion region in the (\invM{\PSGcpmDo}, \invM{\PSGczDo}) plane can be derived, extending to chargino masses of 800\GeV and neutralino masses of 320\GeV.
These are the most stringent limits on this model to date.
For chargino decays into a neutralino and a {\PW} boson, limits on the production cross section have been derived assuming a neutralino mass of 1\GeV, and chargino masses in the range 170--200\GeV have been excluded.

Top squark pair production was also tested, with a focus on compressed decay modes.
A model with the top squark decaying into a top quark and a neutralino was considered.
In the region where $\invM{\PW}<\invM{\stone}-\invM{\PSGczDo}\lesssim \invM{\PQt}$, limits extend up to 420 and 360\GeV for the top squark and neutralino masses, respectively.
An alternative model has also been considered, where the top squark decays into a chargino and a bottom quark, with the chargino subsequently decaying into a {\PW} boson and the lightest neutralino. The mass of the chargino is assumed to be average between the top squark and neutralino masses, which gives a lower bound to the mass difference (\Dm) between the top squark and the neutralino of $\Dm\approx 2\,\invM{\PW}$. This search reduces by about 50\GeV the minimum \Dm excluded in the previous result with two leptons in the final state~\cite{ref:t2tt2lep} from the CMS Collaboration, excluding top squark masses in the range 225--325\GeV for $\Dm\approx 2\,\invM{\PW}$.

In summary, by exploiting the full data set collected by the CMS experiment in 2016, this search extends the existing exclusion limits on the pair production of charginos decaying via sleptons~\cite{ATLASTChipmSlepSnu13TeV}, improving by about 70\GeV the limit on the chargino mass for a massless neutralino. Exclusion limits on top squark pair production extend the results obtained by the CMS Collaboration in final states with two oppositely charged leptons~\cite{ref:t2tt2lep} to the compressed region, where they are competitive with the results obtained by the ATLAS Collaboration in the same decay channel~\cite{ATLASStop}.

\begin{acknowledgments}
We congratulate our colleagues in the CERN accelerator departments for the excellent performance of the LHC and thank the technical and administrative staffs at CERN and at other CMS institutes for their contributions to the success of the CMS effort. In addition, we gratefully acknowledge the computing centers and personnel of the Worldwide LHC Computing Grid for delivering so effectively the computing infrastructure essential to our analyses. Finally, we acknowledge the enduring support for the construction and operation of the LHC and the CMS detector provided by the following funding agencies: BMWFW and FWF (Austria); FNRS and FWO (Belgium); CNPq, CAPES, FAPERJ, and FAPESP (Brazil); MES (Bulgaria); CERN; CAS, MoST, and NSFC (China); COLCIENCIAS (Colombia); MSES and CSF (Croatia); RPF (Cyprus); SENESCYT (Ecuador); MoER, ERC IUT, and ERDF (Estonia); Academy of Finland, MEC, and HIP (Finland); CEA and CNRS/IN2P3 (France); BMBF, DFG, and HGF (Germany); GSRT (Greece); NKFIA (Hungary); DAE and DST (India); IPM (Iran); SFI (Ireland); INFN (Italy); MSIP and NRF (Republic of Korea); LAS (Lithuania); MOE and UM (Malaysia); BUAP, CINVESTAV, CONACYT, LNS, SEP, and UASLP-FAI (Mexico); MBIE (New Zealand); PAEC (Pakistan); MSHE and NSC (Poland); FCT (Portugal); JINR (Dubna); MON, RosAtom, RAS and RFBR (Russia); MESTD (Serbia); SEIDI, CPAN, PCTI and FEDER (Spain); Swiss Funding Agencies (Switzerland); MST (Taipei); ThEPCenter, IPST, STAR, and NSTDA (Thailand); TUBITAK and TAEK (Turkey); NASU and SFFR (Ukraine); STFC (United Kingdom); DOE and NSF (USA).

Individuals have received support from the Marie-Curie program and the European Research Council and Horizon 2020 Grant, contract No. 675440 (European Union); the Leventis Foundation; the A. P. Sloan Foundation; the Alexander von Humboldt Foundation; the Belgian Federal Science Policy Office; the Fonds pour la Formation \`a la Recherche dans l'Industrie et dans l'Agriculture (FRIA-Belgium); the Agentschap voor Innovatie door Wetenschap en Technologie (IWT-Belgium); the Ministry of Education, Youth and Sports (MEYS) of the Czech Republic; the Council of Scientific and Industrial Research, India; the HOMING PLUS program of the Foundation for Polish Science, cofinanced from European Union, Regional Development Fund, the Mobility Plus program of the Ministry of Science and Higher Education, the National Science Center (Poland), contracts Harmonia 2014/14/M/ST2/00428, Opus 2014/13/B/ST2/02543, 2014/15/B/ST2/03998, and 2015/19/B/ST2/02861, Sonata-bis 2012/07/E/ST2/01406; the National Priorities Research Program by Qatar National Research Fund; the Programa Severo Ochoa del Principado de Asturias; the Thalis and Aristeia programs cofinanced by EU-ESF and the Greek NSRF; the Rachadapisek Sompot Fund for Postdoctoral Fellowship, Chulalongkorn University and the Chulalongkorn Academic into Its 2nd Century Project Advancement Project (Thailand); the Welch Foundation, contract C-1845; and the Weston Havens Foundation (USA).
\end{acknowledgments}

\bibliography{auto_generated}
\cleardoublepage \appendix\section{The CMS Collaboration \label{app:collab}}\begin{sloppypar}\hyphenpenalty=5000\widowpenalty=500\clubpenalty=5000\input{SUS-17-010-authorlist.tex}\end{sloppypar}
\end{document}

%% file: SUS-17-010-authorlist.tex
\vskip\cmsinstskip
\textbf{Yerevan Physics Institute, Yerevan, Armenia}\\*[0pt]
A.M.~Sirunyan, A.~Tumasyan
\vskip\cmsinstskip
\textbf{Institut f\"{u}r Hochenergiephysik, Wien, Austria}\\*[0pt]
W.~Adam, F.~Ambrogi, E.~Asilar, T.~Bergauer, J.~Brandstetter, M.~Dragicevic, J.~Er\"{o}, A.~Escalante~Del~Valle, M.~Flechl, R.~Fr\"{u}hwirth\cmsAuthorMark{1}, V.M.~Ghete, J.~Hrubec, M.~Jeitler\cmsAuthorMark{1}, N.~Krammer, I.~Kr\"{a}tschmer, D.~Liko, T.~Madlener, I.~Mikulec, N.~Rad, H.~Rohringer, J.~Schieck\cmsAuthorMark{1}, R.~Sch\"{o}fbeck, M.~Spanring, D.~Spitzbart, A.~Taurok, W.~Waltenberger, J.~Wittmann, C.-E.~Wulz\cmsAuthorMark{1}, M.~Zarucki
% ...
\vskip\cmsinstskip
% ...
1:  Also at Vienna University of Technology, Vienna, Austria\\
% ...
